\documentclass{article}
\usepackage{booktabs,array}
\usepackage{graphicx}
\usepackage{amsmath}

\usepackage{amssymb}
\usepackage{xcolor}
\usepackage[left=3.00cm, right=3.00cm, top=2.00cm, bottom=2.00cm]{geometry}
\usepackage{authblk}

\title{Two simple textures of the magic neutrino mass matrix}
\author[1,2]{Kanwaljeet S. Channey \footnote{kjschanney@outlook.com}}
\author[1]{Sanjeev Kumar\footnote{sanjeevkumarverma@outlook.in}}
\affil[1]{Department of Physics and Astrophysics, University of Delhi, Delhi, 110007, India.}
\affil[2]{Department of Physics, University Institute of Sciences, Chandigarh University, Mohali, Punjab 140413, India.}  
\date{}
\begin{document}
\maketitle
\begin{abstract}
The Tri-Bimaximal (TBM) mixing predicts a vanishing $\theta_{13}$. This can be
  attributed to the inherited $\mu-\tau$ symmetry of TBM mixing. We break its
  $\mu-\tau$ symmetry by adding a complex magic matrix with one variable to TBM neutrino mass matrix
  with one vanishing eigenvalue. We present two such textures and study their
  phenomenological implications.
\end{abstract}

In the Standard Model of electroweak and strong interactions, neutrino flavor
states $\nu_{l}$ $(l=e,\mu,\tau)$ are the states which
form weak doublets with the corresponding charged lepton states $l$:
\begin{equation}
  \label{eq:chargedcurrent}
  j^{\mu} = \bar{l} \gamma^{\mu}(1-\gamma_5) \nu_l.
\end{equation}
These neutrino states $\nu_l$ are the coherent combinations of the neutrino mass
states $\nu_i$ $(i=1,2,3)$ which are also the eigenstates of the Hamiltonian in
vacuum. Neutrino flavor and mass states are related by the relation 
\begin{equation}
  \label{eq:superposition}
  \nu_l = U_{\text{PMNS}} \nu_i,
\end{equation}
where $U_{\text{PMNS}}$ is the unitary matrix called
Pontecorvo-Maki-Nakagawa-Sakata mixing matrix. By convention, PMNS mixing matrix
is defined as

\begin{equation}
 \label{eq:pmns}
U_{\text{PMNS}} = \left(\begin{array}{ccc}
                   c_{12} c_{13} & s_{12} c_{13} & s_{13} e^{-\iota \delta} \\
                   -s_{12}c_{23}-c_{12} s_{23} s_{13} e^{\iota \delta} & c_{12} c_{23}-s_{12} s_{23} s_{13} e^{\iota \delta} & s_{23} c_{13} \\
                   s_{12} s_{23}-c_{12} c_{23} s_{13} e^{\iota \delta} & -c_{12} s_{23}-s_{12} c_{23} s_{13} e^{\iota \delta} & c_{23} c_{13} \\
                 \end{array}
               \right)
             \end{equation}         
             where $s_{ij}=\sin \theta_{ij}$, $c_{ij}=\cos \theta_{ij}$ $(i,j=1,2,3)$ and
             $\delta$ is the Dirac type CP violating phase.
             
In the flavor basis, where lepton mass matrix is diagonal, the neutrino mass
matrix $M_{\nu}$ is related with the unitary mixing matrix and the complex
neutrino masses $m_d=\text{diag}(m_1,m_2 e^{2 \iota \alpha},m_3 e^{2 \iota
  \beta})$ by the relation

\begin{equation}
               \label{eq:diagonaliztion}
M_{\nu} = U^*_{\text{PMNS}} m_d U^\dagger_{\text{PMNS}},
\end{equation}
where $\alpha$ and $\beta$ are the Majorana phases.
While analyzing the experimental results on neutrino mixing and mass matrices,
we can often look for some particular features like equalities \cite{Dev2013,Han2017},
zeros \cite{Frampton2002,Grimus2004,Merle2006,Dev2007b,Dev2007d,Fritzsch2011,Grimus2013,Blankenburg2013,Liao2013a,Dziewit2017}, hybrids of zeros
and equalities \cite{Kaneko2005,Goswami2010,Dev2010,Dev2010a,Kalita2016}, zero trace \cite{He2003a}, or some other pattern \cite{Lavoura2005,Lashin2008,Lashin2009,Wang2013a} in their elements.

Harrison, Perkins and Scott proposed one such type of mixing
matrix \cite{Harrison2002167} which had $\nu_2$ trimaximally mixed and $\nu_3$
bimaximally mixed. Hence, they named it Tri-Bimaximal
(TBM) mixing. The TBM mixing matrix is 
\begin{equation}
  \label{eq:utbm}
  U_{\text{TBM}} = \left(\begin{array}{ccc}
    \sqrt{\frac{2}{3}} & \frac{1}{\sqrt{3}}& 0 \\
    -\sqrt{\frac{1}{6}} & \frac{1}{\sqrt{3}}& \frac{1}{\sqrt{2}} \\
     -   \sqrt{\frac{1}{6}} & \frac{1}{\sqrt{3}} & -\frac{1}{\sqrt{2}} 
  \end{array}\right).
\end{equation}
The mass matrix $M_{\text{TBM}}$ corresponding to $U_{\text{TBM}}$
can be written by using \mbox{Eq. (\ref{eq:diagonaliztion})}
\begin{equation}
  \label{eq:mtbm}
       M_{TBM} = \left( \begin{array}{ccc}
                          a & b & b \\
                          b &a+ d & b-d \\
                          b & b-d & a+d \\
                        \end{array} \right).
\end{equation}

TBM mixing matrix predicts $\sin^2 \theta_{13} = 0$, $\sin^2
\theta_{12}=\frac{1}{3}$ and $\sin^2 \theta_{23} = \frac{1}{2}$. Mixing angles $\theta_{12}$ and
$\theta_{23}$ are in agreement at 3$\sigma$ with their experimental values, $\sin^2 \theta_{12} =
0.306^{+0.012}_{-0.012}$ and $\sin^2 \theta_{23} = 0.441^{+0.027}_{-0.021}$,
provided by the latest global fit of the neutrino experimental
data \cite{Esteban2016b}.

Another parametrization of the mixing matrix was given in Ref.
  \cite{King2008244}. In this parametrization, the mixing matrix was expressed
  as expansion in powers of the deviations of reactor, solar and atmospheric
mixing angles from their TBM value . Let $r,s,$ and
$a$ are the real parameters which give deviations to the reactor, solar and
atmospheric mixing angles from their TBM values:
\begin{equation}
\label{eq:kingangles}
\sin \theta_{13} = \frac{r}{\sqrt{2}}, \hspace{0.5cm}
\sin \theta_{12} = \frac{1}{\sqrt{3}}(1+s), \hspace{0.5cm}
\sin \theta_{23} = \frac{1}{\sqrt{2}}(1+a).
\end{equation}
Since the parameters $r,s,$ and $a$ are very small, we can expand the mixing
matrix about $U_\text{TBM}$ in the powers of $r,s,$ and $a$. We present here the mixing
matrix to the first order in $r,s,$ and $a$
\begin{equation}
\label{eq:9}
U_p \approx \left(
\begin{array}{ccc}
  \sqrt{\frac{2}{3}}\left( 1- \frac{1}{2}s \right) & \frac{1}{\sqrt{3}}(1+s)&\frac{1}{\sqrt{2}}re^{\iota \delta}\\
  - \frac{1}{\sqrt{6}}(1+s-a+re^{\iota \delta}) & \frac{1}{\sqrt{3}}(1-\frac{1}{2}s-a-\frac{1}{2}re^{\iota \delta}) & \frac{1}{\sqrt{2}}(1+a)\\
  \frac{1}{\sqrt{6}}(1+s+a-re^{\iota \delta})&-\frac{1}{\sqrt{3}}(1-\frac{1}{2}s+a-\frac{1}{2}re^{\iota \delta})& \frac{1}{\sqrt{2}}(1-a)
\end{array}
\right).
\end{equation}

The mixing angle $\theta_{13}$ is non zero as measured by the
recent experiments: T2K\cite{T2K_NON_ZERO_Theta13_PhysRevLett.107.041801}, Daya Bay\cite{Daya_Bay_reactor_angle_PhysRevLett.108.171803}, RENO\cite{RENO_THETA13_PhysRevLett.108.191802}
and DOUBLE CHOOZ\cite{Double_CHOOZ_reactor_angle_PhysRevLett.108.131801}. This leads to the realization that although
TBM ansatz is ruled out by the experiments, it can still be used
as leading order contribution to the neutrino mass matrix. We can add
perturbations to $M_{\text{TBM}}$ so as to generate the non-zero $\theta_{13}$. TBM mass
matrix obeys both the magic symmetry and the $\mu-\tau$ exchange symmetry. Magic symmetry
means sum of the elements of each row and column of mass matrix remains the
same, whereas $\mu-\tau$ exchange symmetry means that the neutrino mass matrix is invariant under the simultaneous
interchange of its second and third ($\mu$-$\tau$) indices.

A neutrino mass matrix that is invariant under magic symmetry and $\mu-\tau$
exchange symmetry predicts maximal $\theta_{23}$ and vanishing $\theta_{13}$. These predictions
are very close to the present neutrino oscillation data. This indicates that
we can satisfy the present experimental data by introducing small
perturbations to the magic mass matrix. Magic symmetry also provides sum rules
between the mixing angles due to trimaximal structure of $\nu_2$ \cite{Lam2006} which in return reduces the number of free parameters. These sum rules can be tested at the future neutrino oscillation experiments.

In the present paper, we propose two simple textures of $M_{\text{magic}}$ that break the
$\mu-\tau$ symmetry of TBM neutrino mass matrix but preserve its magic
symmetry. These textures can be written as
\begin{equation}
  \label{eq:1}
  M^i_{\text{magic}} = M_{\text{TBM}} +M'_i ,\hspace{0.5cm} (i=a,b)
\end{equation}
The $\mu-\tau$ breaking term $M_i$ in these textures is function of only one complex variable $\eta = z
e^{\iota \chi}$. To reduce the number of independent variables, in our study we
have considered the $M_{\text{TBM}}$ with
vanishing lowest eigenvalue. This assumption will lead to the condition that
$b=a$ in the Eq. \ref{eq:mtbm}. Forms of $M_{\text{TBM}}$ and $M'_i$ $(i=a,b)$
studied in the present work for normal hierarchy are:
\begin{equation}
  \label{eq:textures}
M_{\text{TBM}} = \left(
\begin{array}{ccc}
 a & a & a \\
 a & a+d & a-d \\
 a & a-d & a+d \\
\end{array}
\right), M'_a = \left(
\begin{array}{ccc}
 0 & 0 & \eta  \\
 0 & 0 & \eta  \\
 \eta  & \eta  & -\eta  \\
\end{array}
\right), M'_b = \left(
\begin{array}{ccc}
 0 & \eta  & 0 \\
 \eta  & 0 & 0 \\
 0 & 0 & \eta  \\
\end{array}
\right).
\end{equation}
We then study the phenomenological implications for these textures of neutrino mass matrix.

While perturbing the TBM mass matrix by adding an extra matrix, we can
break $M_{\text{TBM}}$ in such a way that out of the two symmetries that it possesses, we break only one.
Since $\mu-\tau$ symmetry predicts vanishing $\theta_{13}$, preserving magic
symmetry is a feasible choice. 

If a transformation $G_j$ of the neutrino fields leaves the neutrino
mass matrix unchanged such that
\begin{equation}
  \label{eq:transformation}
G^T_j M_{\nu} G_j = M_{\nu},
\end{equation}
 the transformation $G_j$ is called a symmetry of mass matrix $M_{\nu}$. The transformation matrix
can be calculated using the relation, $G_j = 1-u_j u^T_j$ $(j=1,2,3)$ where $u_j$ is the column
of matrix corresponding to the symmetry $G_j$. The transformation matrix corresponding to the magic symmetry
is given below
\begin{equation}
  \label{eq:magictransformation}
G_2 = \left(
\begin{array}{ccc}
 \frac{1}{3} & -\frac{2}{3} & -\frac{2}{3} \\
 -\frac{2}{3} & \frac{1}{3} & -\frac{2}{3} \\
 -\frac{2}{3} & -\frac{2}{3} & \frac{1}{3} \\
\end{array}
\right).
\end{equation}
Therefore, a mass matrix $M_{\text{magic}}$ that preserves the magic symmetry will obey the relation
$G^T_2 M_{\text{magic}} G_2 = M_{\text{magic}}$. Mixing matrix corresponding to such mass matrices will have
their middle column same as that of $U_{\text{TBM}}$ (trimaximal) and can be described in
terms of two independent variables $\theta$ and $\phi$
\begin{equation}
  \label{eq:tm2}
  U_{\text{TM}}=\left(
    \begin{array}{ccc}
      \sqrt{\frac{2}{3}}
      \cos  \theta   &  \frac{1}{\sqrt{3}}  & \sqrt{\frac{2}{3}}
                                              \sin  \theta   \\
      \frac{e^{i \phi } \sin  \theta  -\frac{\cos  \theta }{\sqrt{3}}}{\sqrt{2}}&  \frac{1}{\sqrt{3}}  & \frac{-e^{i
                                                                                                         \phi } \cos  \theta  -\frac{\sin
                                                                                                         \theta  }{\sqrt{3}}}{\sqrt{2}} \\
      \frac{-\frac{\cos
      \theta  }{\sqrt{3}}-e^{i \phi } \sin
      \theta  }{\sqrt{2}}&                 \frac{1}{\sqrt{3}}   & \frac{e^{i
                                                                  \phi } \cos  \theta  -\frac{\sin
                                                           \theta  }{\sqrt{3}}}{\sqrt{2}} \\
    \end{array}
  \right).
\end{equation}
This mixing matrix, has a trimaximally mixed column leading to its nomenclature
as trimaximal mixing. One of the general form for $M_{\text{magic}}$ can be written as
\begin{equation}
  \label{eq:tm2mass}
  M_{\text{magic}} =   \left(
\begin{array}{ccc}
 a & b & c \\
 b & a+d & c-d \\
 c & c-d & a+b-c+d \\
\end{array}
\right).
\end{equation}
This mass matrix can be diagonalized by using the equation
\begin{equation}
  \label{eq:diag}
M_d = U^T_{\text{TM}}M_{\text{magic}}U_{\text{TM}}.
\end{equation}
The rationale of choosing the form of $M_{\text{magic}}$ as given in Eq.
(\ref{eq:diag}) is that it reduces to $M_{\text{TBM}}$ (Eq.
(\ref{eq:mtbm})) for $c=b$. It is the difference of $b$ and $c$ that breaks the
$\mu-\tau$ symmetry of $M_{\text{TBM}}$. Therefore, to break $\mu-\tau$ exchange
symmetry and to generate non-zero $\theta_{13}$, we can allow $b$ and $c$ to
differ by a small \mbox{amount ($\eta$).}

The diagonal elements of $M_d$ will give us neutrino masses and Majorana phases, whereas the off-diagonal
elements, when equated to zero, will give the variables $\theta$ and $\phi$ of
$U_{\text{TM}}$ in terms of the parameters of $M_{\text{magic}}$. We can
calculate the mixing angles in terms of $\theta$ and $\phi$ from the elements of
$U=U_{\text{TM}}$ using the relations
\begin{equation}
  \label{eq:anglesgenform}
\sin^2 \theta_{12} = \frac{|U_{12}|^2}{1-|U_{13}|^2}, \hspace{0.5cm}  \sin^2 \theta_{23} = \frac{|U_{23}|^2}{1-|U_{13}|^2},\hspace{0.5cm}  \sin^2 _{13} = |U_{13}|^2.
  \end{equation}
  We can calculate the Dirac phase $\delta$ from the Jarlskog rephasing
  invariant measure of CP violation,
\begin{equation}
  \label{eq:jarlskog}
J = Im[U_{12} U_{23} U^*_{13} U^*_{22}],
\end{equation}
using the relation,
\begin{equation}
  \label{eq:2jarlskog2}
  J = c_{12} s_{12} c_{23} s_{23} c^2_{13} s_{13} \sin \delta.
\end{equation}
Neutrino masses and Majorana phases can be calculated from $M_d$ using the following relations
  \begin{equation}
    \label{eq:masses}
    |m_1| =|[M_d]_{11}|, \hspace{0.5cm}|m_2| = |[M_d]_{22}|,\hspace{0.5cm}|m_3|=|[M_d]_{33}|,
  \end{equation}
  \begin{equation}
    \label{eq:phases}
\alpha = \frac{1}{2} \arg \left(\frac{[M_d]_{22}}{[M_d]_{11}} \right)  , \hspace{0.5cm} \beta = \frac{1}{2} \arg \left(\frac{[M_d]_{33}}{[M_d]_{11}} \right).
  \end{equation}
We can write $M_{\text{magic}}$ as the sum of $M_{\text{TBM}}$ and a $\mu-\tau$
symmetry breaking term $M'$:
\begin{equation}
  \label{eq:symbr}
  M_{\text{magic}} = M_{\text{TBM}} + M'
\end{equation}
where $M'$ is also invariant under $G_2$.
Looking at Eq. (\ref{eq:tm2mass}), we observe that we can write $M_{\text{magic}}$ as sum of four matrices:

\begin{equation}
  \label{eq:linearcombination}
  M_{\text{magic}}=\left(
\begin{array}{ccc}
 a & 0 & 0 \\
 0 & 0 & a \\
 0 & a & 0 \\
\end{array}
\right)+\left(
\begin{array}{ccc}
 0 & 0 & 0 \\
 0 & d & -d \\
 0 & -d & d \\
\end{array}
\right)+\left(
\begin{array}{ccc}
 0 & b & 0 \\
 b & 0 & 0 \\
 0 & 0 & b \\
\end{array}
\right)+\left(
\begin{array}{ccc}
 0 & 0 & c \\
 0 & 0 & c \\
 c & c & -c \\
\end{array}
\right).
\end{equation}
\begin{figure}[h]
  \centering
 \includegraphics[width=0.49\textwidth]{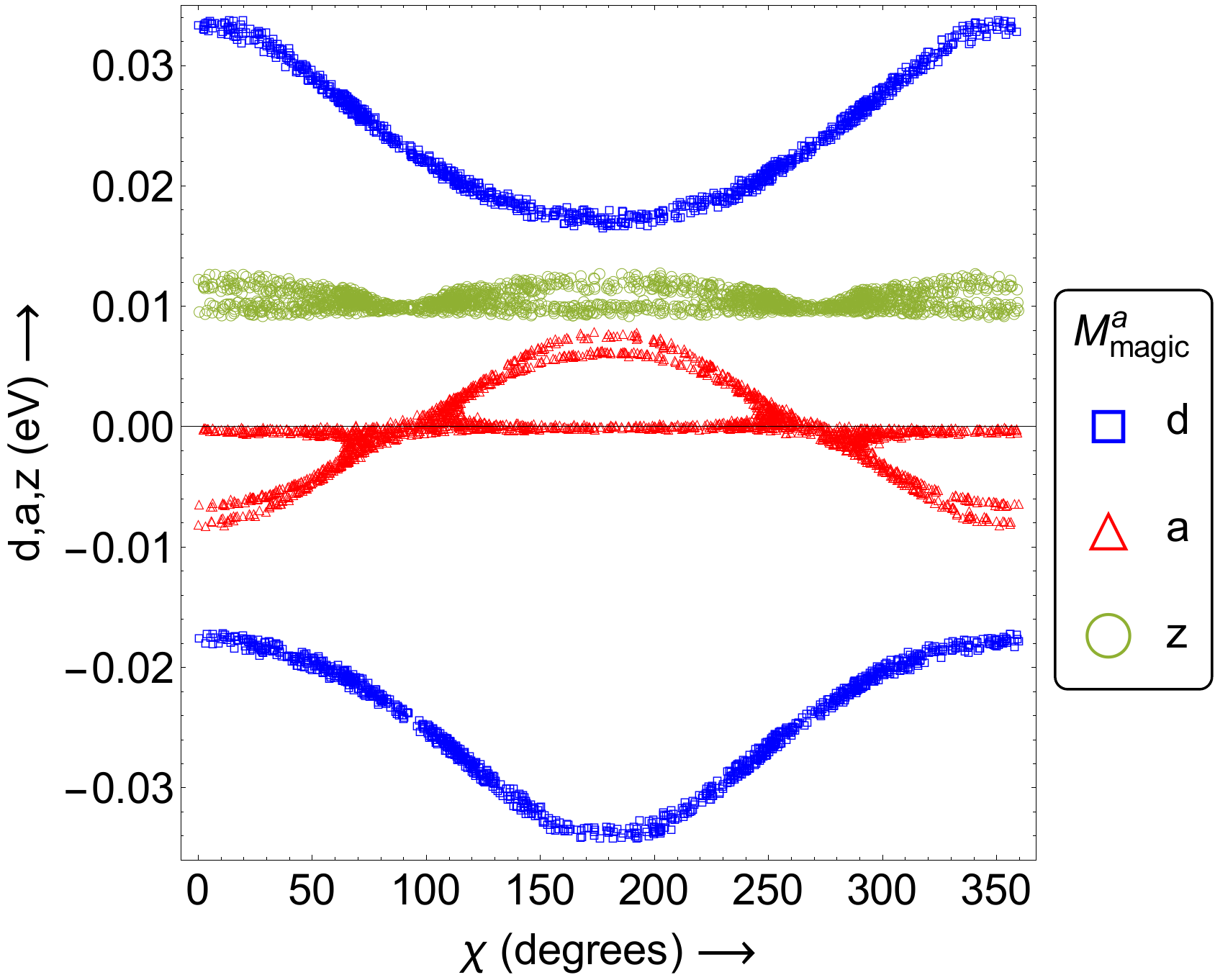} \includegraphics[width=0.49\textwidth]{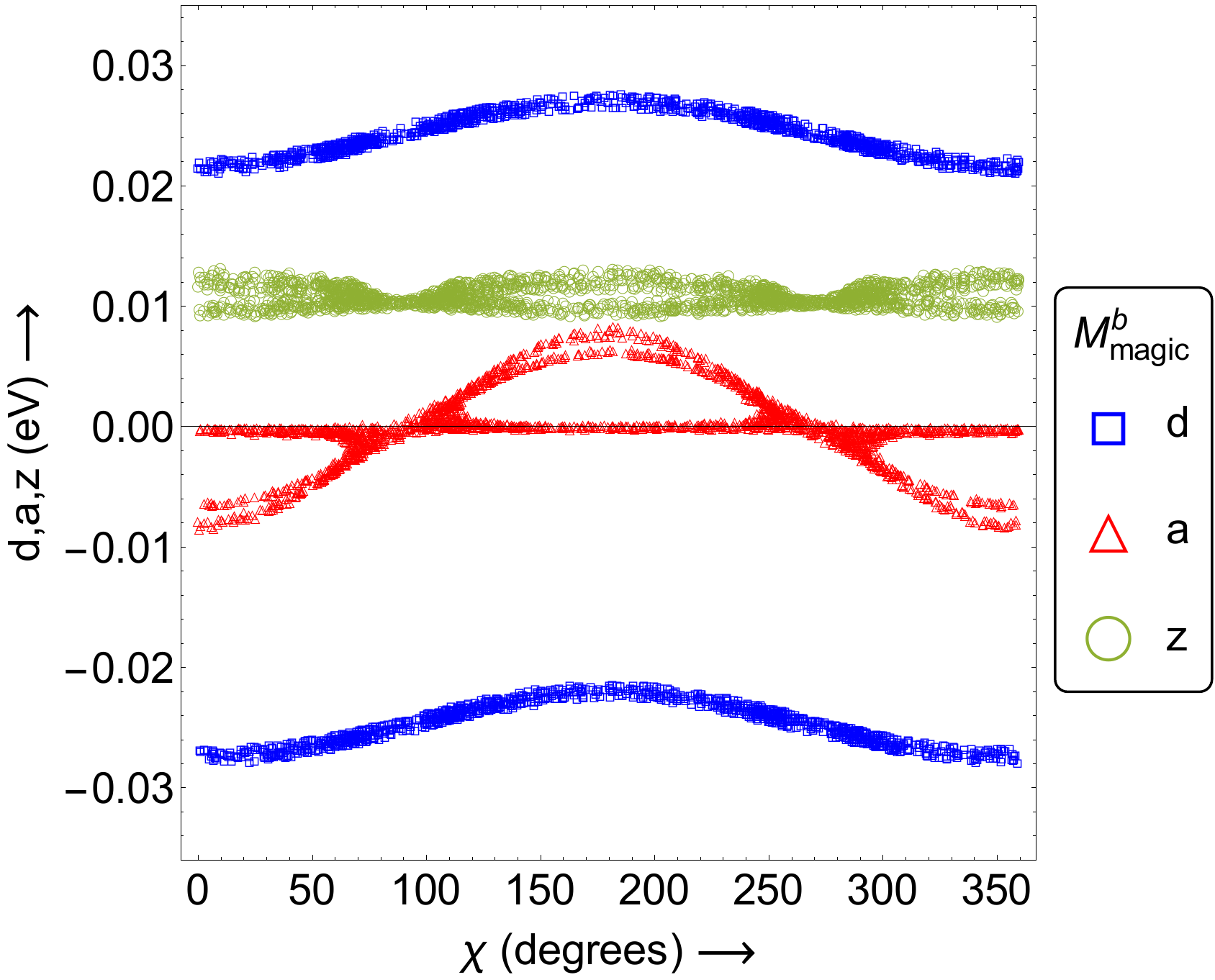}
    \caption{The allowed parameter space for $a, b, z,$ and $\chi$ for the textures $M^a_\text{magic}$
      and $M^b_\text{magic}$.}
 \label{fig:massvschi}
\end{figure}
There are two ways to write $M_{\text{magic}}$ as combination of
$M_{\text{TBM}}$ and $M'$ as shown in Eq.
(\ref{eq:symbr}). First is by considering $b=a$, $c=a+\eta$
\begin{equation}
  \label{eq:magic1}
  M^a_{\text{magic}} = \left(
\begin{array}{ccc}
 a & a & a \\
 a & a+d & a-d \\
 a & a-d & a+d \\
\end{array}
\right) + \left(
\begin{array}{ccc}
  0 & 0 & \eta \\
 0 & 0 & \eta \\
 \eta & \eta & -\eta \\
\end{array}
\right)
\end{equation}
and second is by
considering $b=a+\eta$, $c=a$
\begin{equation}
  \label{eq:magic2}
  M^b_{\text{magic}} = \left(
\begin{array}{ccc}
 a & a & a \\
 a & a+d & a-d \\
 a & a-d & a+d \\
\end{array}
\right) +\left(
\begin{array}{ccc}
 0 & \eta & 0 \\
 \eta & 0 & 0 \\
 0 & 0 & \eta \\
\end{array}
\right),
\end{equation}
where $\eta = z e^{\iota \chi}$ is responsible for the breaking of $\mu-\tau$ symmetry. The assumption $a=b$ in $M_{\text{TBM}}$ results in vanishing lowest
eigen value of $M_{\text{TBM}}$. We made this assumption to reduce the number of
free parameters in $M_{\text{TBM}}$. This gives us the two textures studied in
this paper given in Eq. (\ref{eq:textures}) ($M^a_\text{magic}$ and $M^b_\text{magic}$).

We can diagonalize these mass matrices by using Eq. (\ref{eq:diag}) and obtain
our predictions from Eqs. (\ref{eq:anglesgenform}-\ref{eq:phases}). Equating the
nondiagonal entry $[m_d]_{13}$ with zero for these textures will give us
predictions for the variables
$\theta$ and $\phi$ in terms of $a$, $d$, $z$ and $\chi$.

\begin{figure}[h]
  \centering
\includegraphics[width=0.49\textwidth]{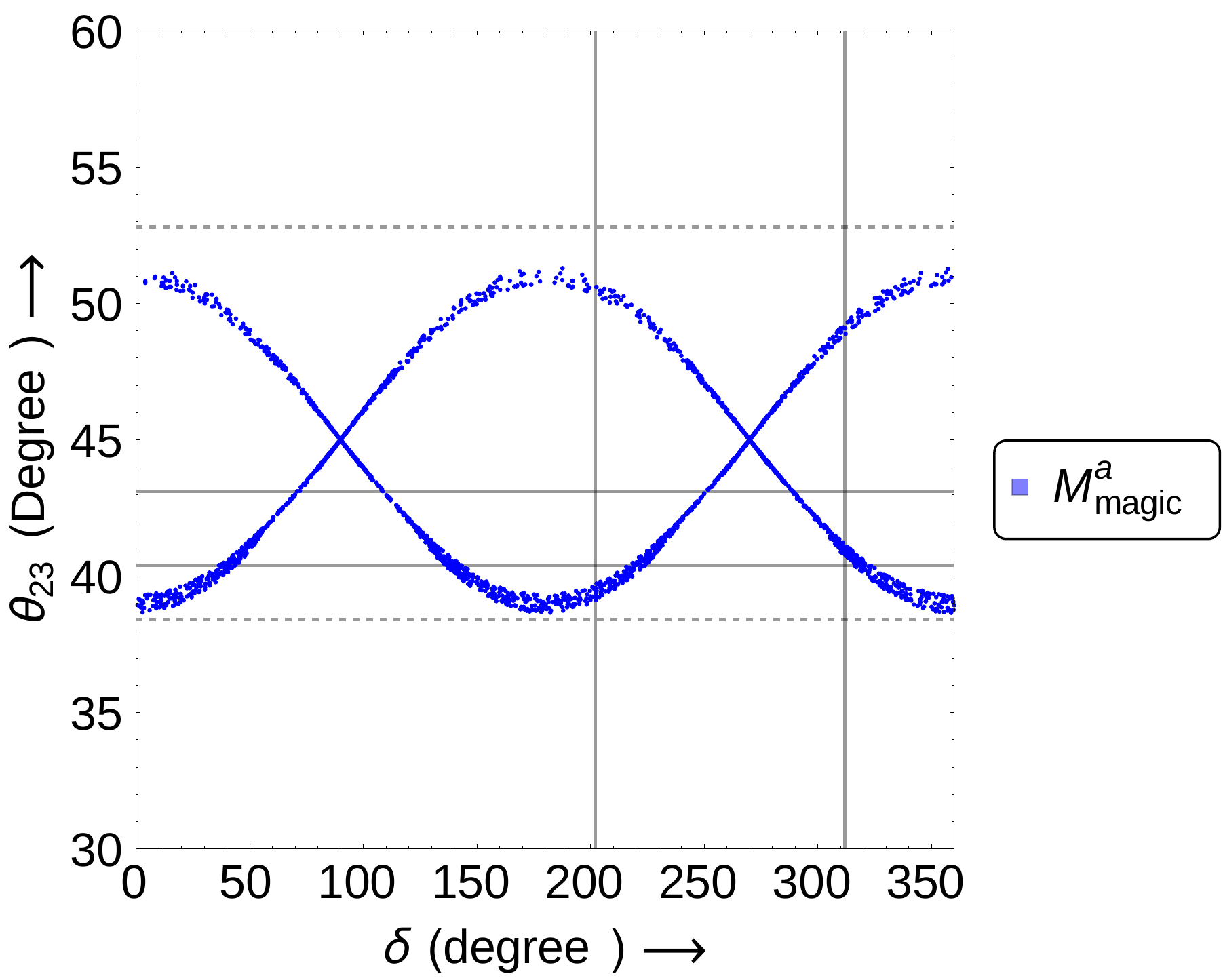} \includegraphics[width=0.49\textwidth]{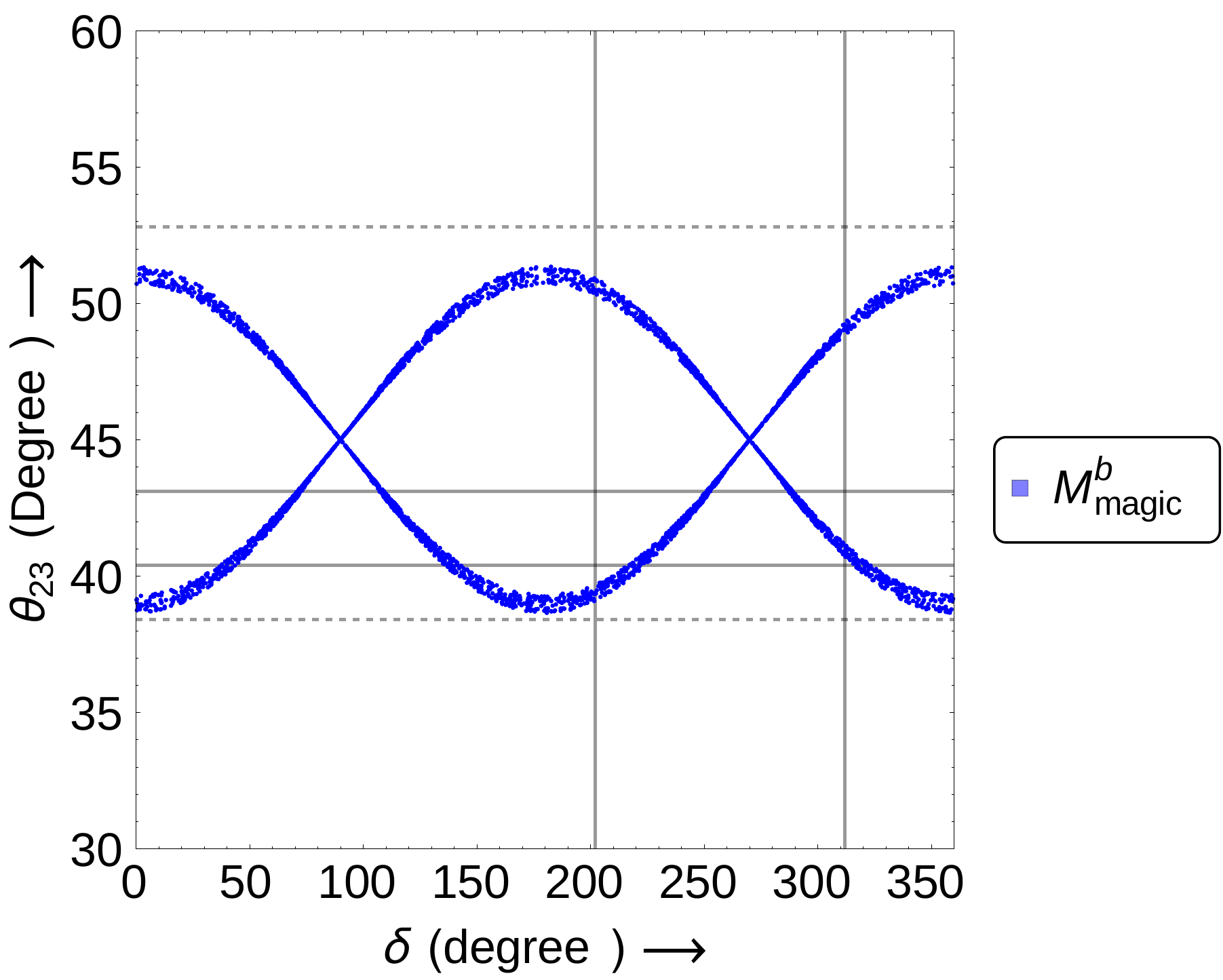}  
    \caption{The correlations between atmospheric angle $\theta_{23}$ and CP
      violating phase $\delta$ for both
      the textures $M^a_{\text{magic}}$ and $M^b_{\text{magic}}$. Here
      dashed lines represent the  3$\sigma$ experimental range and solid lines represent the 
       1$\sigma$ experimental range.}
 \label{fig:th23delta}
\end{figure}
For texture $M^a_{\text{magic}}$, the nondiagonal entry $[m_d]_{13}$, $\theta$
and $\phi$ are as follows:
\begin{align}
   \label{eq:theta phi for a}
   [m_d]_{13} &= \frac{1}{4} e^{-2 i \phi } \left(2 \sqrt{3} z \cos 2 \theta  e^{i
       (\chi +\phi )} \right) \nonumber \\ &- \frac{1}{4} e^{-2 i \phi }\left( \sin 2 \theta  \left(4 d +e^{i \chi } z \left(-3+e^{2 i \phi }\right)\right)\right), \\
 \tan \phi &= -\frac{d \tan \chi }{d-z \sec \chi }, \\
\tan 2 \theta &= -\frac{2 \sqrt{3} z \cos (\phi -\chi )}{-4 d \cos 2 \phi -z (\cos \chi -3 \cos (2 \phi -\chi ))}.
\end{align}
 For texture $M^b_{\text{magic}}$ we obtain the following relations for $[m_d]_{13}$, $\theta$
 and $\phi$:
\begin{align}
  \label{eq:theta phi for b}
[m_d]_{13} &= \frac{1}{4} e^{-2 i \phi } \left(-\sin 2 \theta  \left(4 d+e^{i
                \chi } z \left(1+e^{2 i \phi }\right)\right)\right) \\ &- \frac{1}{4} e^{-2 i \phi } \left(2 \sqrt{3} z 
  \cos 2 \theta  e^{i (\chi +\phi )}\right), \\
   \phi &= -\chi, \\
   \tan 2 \theta &= \frac{\sqrt{3} z \cos (\phi -\chi )}{-2 d \cos 2 \phi -z \cos \phi  \cos (\phi -\chi )}.
 \end{align}
\begin{table}[b]
  \caption{Experimental values of the oscillation parameters.
    Experimental bounds of $\theta_{23}$ are not used in the
    present Monte Carlo analysis.}
  \centering
$\begin{array}{cc}
     \hline
     \text{ Parameters} & \text{ 3$\sigma$ range} \\
          \hline \\
     
     \Delta m^2_{12}/ (10^{-5}) eV^2 &  7.03 \rightarrow 8.09 \\
     
     \Delta m^2_{23}/ (10^{-3}) eV^2 & 2.407 \rightarrow 2.643 \\
     
     \theta_{13}/ ^o & 7.99 \rightarrow 8.90 \\
     
     \theta_{12}/ ^o &  31.38 \rightarrow 35.99 \\

     \theta_{23}/ ^o &  38.4 \rightarrow 52.8 \\
   \hline
   \end{array}$
  \label{tab:experimental}
\end{table}
Corresponding mixing angles and Dirac type CP violating phase then can be
calculated from these $\theta$ and $\phi$ by using the following relations
\begin{eqnarray}
  \sin^2 \theta_{12} &=& \frac{1}{3-2 \sin ^2 \theta },   \label{eq:tm2angles}\\
  \sin^2 \theta_{23} &=& \frac{1}{2} \left(1+ \frac{\sqrt{3} \sin 2 \theta  \cos
                           \phi }{3-2 \sin ^2 \theta }\right),   \label{eq:tm2angles2}\\
  \sin^2 \theta_{13} &=& \frac{2 \sin ^2 \theta }{3},\\
   \csc^2 \delta &=& \csc ^2 \phi -\frac{3 \sin ^2 2 \theta  \cot ^2 \phi
                         }{\left(3-2 \sin ^2 \theta \right)^2}.  \label{eq:tm2angles3}
\end{eqnarray}

\begin{figure}[h]
\centering
\includegraphics[width=0.49\textwidth]{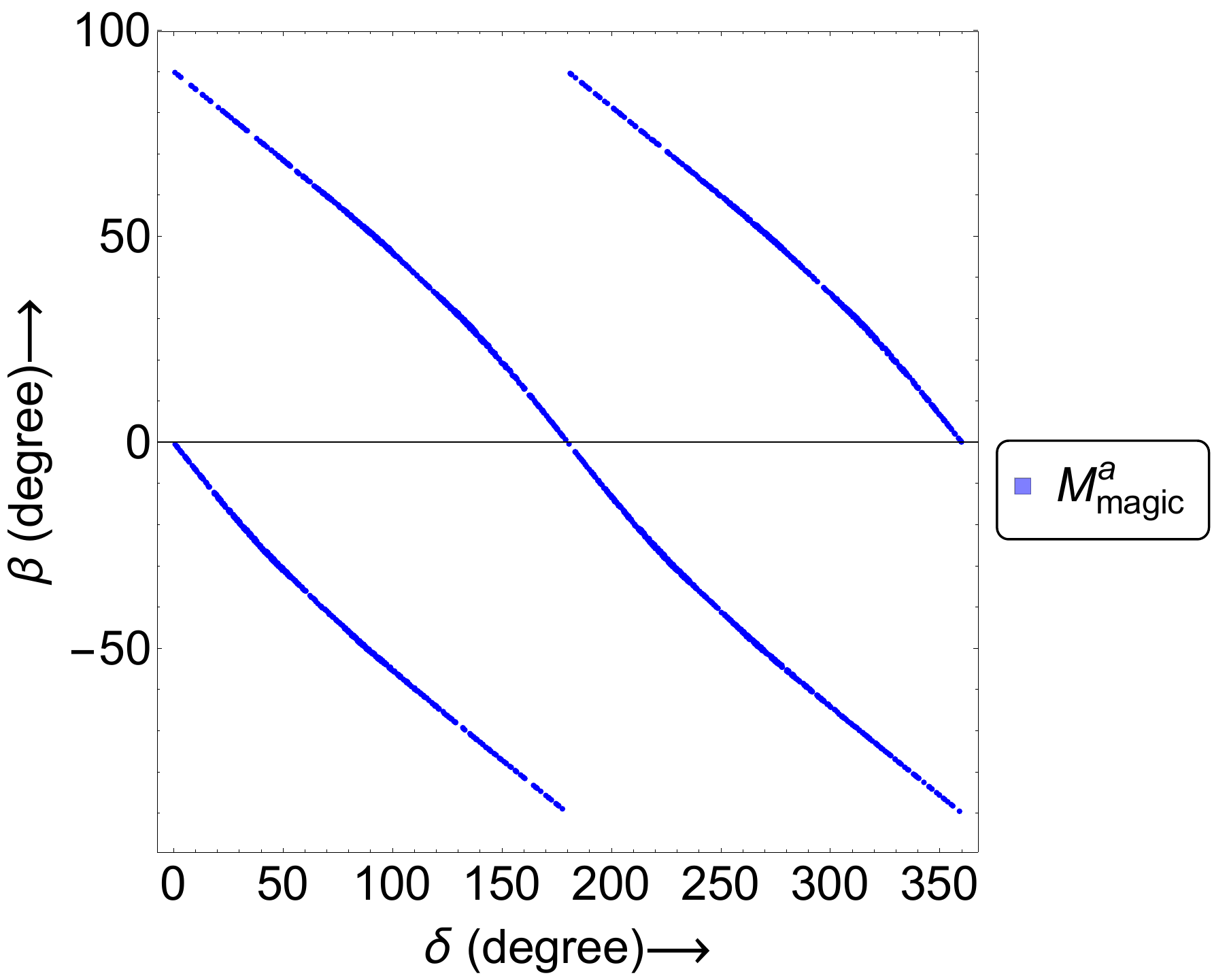} \includegraphics[width=0.49\textwidth]{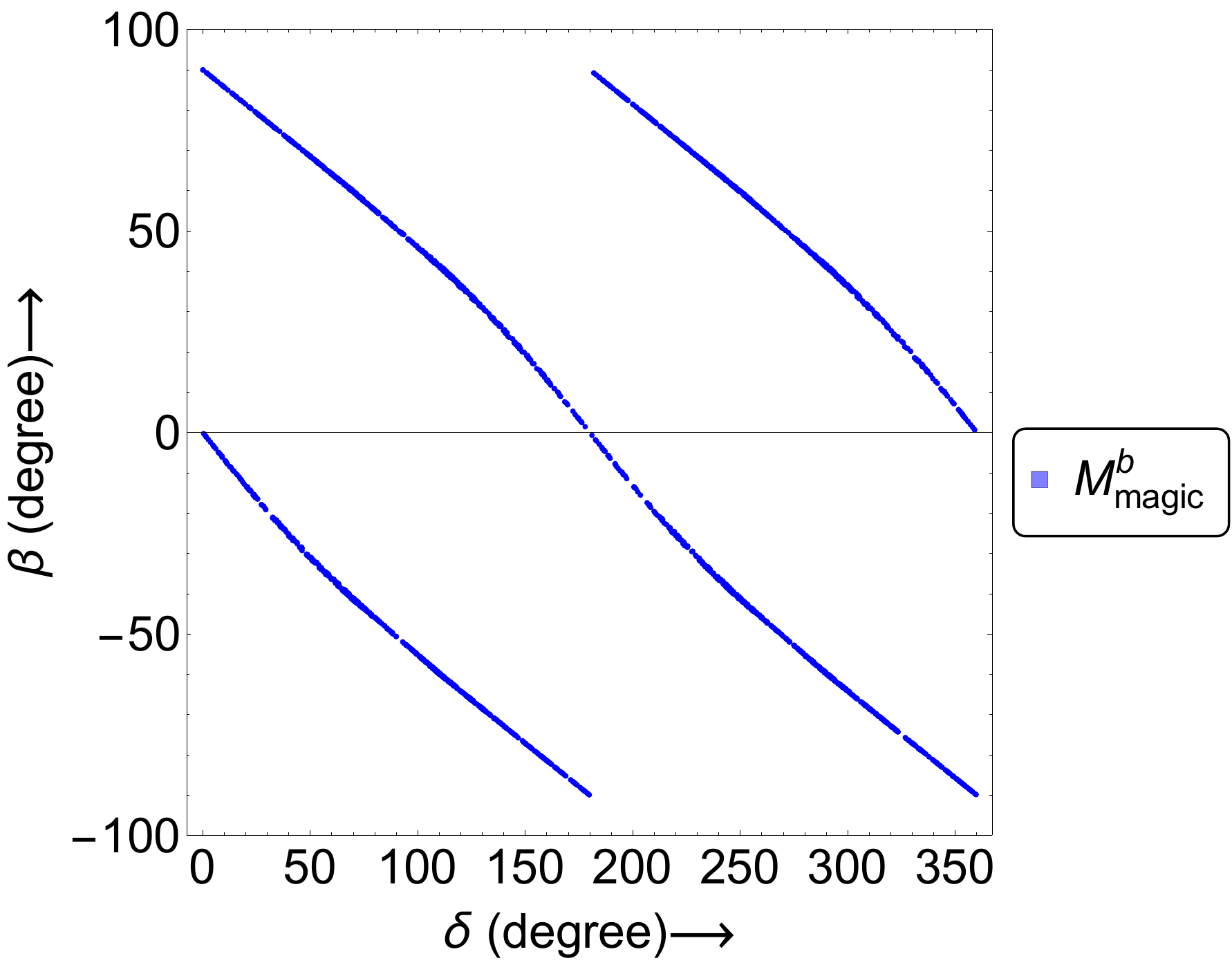}  
\caption{Variation of $\beta$ with $\delta$ for both
      the textures $M^a_{\text{magic}}$ and $M^b_{\text{magic}}$.}
 \label{fig:betadelta}
\end{figure}
\begin{table}[b]
\centering
  \caption{Allowed ranges of the parameters of mass matrix.}
\begin{tabular}{ccc}
    \toprule
    Parameters & \multicolumn{2}{c}{Allowed 3$\sigma$ range} \\
    \hline
                   & $M^a_{\text{magic}}$ & $M^b_{\text{magic}}$ \\
    \cmidrule(lr){2-2}\cmidrule(lr){3-3} \\
    a & [-0.008,0.008] & [-0.008,0.008] \\
    d & [0.016,0.034]$\cup$[-0.034,-0.016] & [0.021,0.027] $\cup$ [-0.027,-0.021] \\
    z & [0.009,0.013] & [0.009,0.013]\\
    \bottomrule
\end{tabular}
    \label{tab:AllowedRanges}
  \end{table}
  Substituting the values of $\theta$ and $\phi$ in terms of $a$, $b$, $z$ and
  $\chi$ in $[m_d]_{11}$, $[m_d]_{22}$ and $[m_d]_{33}$, we can obtain the three
  neutrino masses ($m_1$, $m_2$ and $m_3$) and the Majorana phases ($\alpha$ and
  $\beta$) using Eqs. (\ref{eq:masses}, \ref{eq:phases}), where

\begin{align}
  \label{eq:mass elements}
 \left[ m_d \right]_{11} &= -\frac{1}{2} e^{-2 i \phi } \left(\sin^2
               \theta \left(-4 d+3 e^{i \chi } z\right) \right) \\ &- \frac{1}{2} e^{-2 i \phi } \left( \left( 2 \sqrt{3} z \cos
                             \theta  e^{i (\chi +\phi )}\right)
                             + z \cos ^2 \theta  e^{i (\chi +2 \phi )}\right),\\
\left[ m_d \right]_{22}  &= 3 a + e^{i \chi } z,
\end{align}
and
\begin{align}
 \left[ m_d \right]_{33}   &= \frac{1}{2} e^{-2 i \phi } \left(\cos ^2 \theta 
                               \left(4 d-3 e^{i \chi } z\right) \right)\\ &+ \frac{1}{2} e^{-2 i \phi } \left( z \sin \theta 
                               e^{i (\chi +\phi )} \left(2 \sqrt{3} \cos \theta
                               -e^{i \phi } \sin \theta \right)\right).
\end{align}
Similarly, for texture $M^b_{\text{magic}}$, we have
\begin{align}
  \label{eq:3mass elements 2}
  \left[ m_d \right]_{11}  &= \frac{1}{2} e^{-2 i \phi } \left(\sin ^2 \theta 
                               \left(4 d+e^{i \chi } z\right)+\sqrt{3} z \sin 2
                               \theta  e^{i (\chi +\phi )} \right)\\ &+ \frac{1}{2} e^{-2 i \phi } \left( z \cos ^2 \theta 
                               \left(-e^{i (\chi +2 \phi )}\right)\right),\\
  \left[ m_d \right]_{22}  &= 3 a+e^{i \chi } z ,
\end{align}
and
\begin{align}                                 
   \left[ m_d \right]_{33}  &= \frac{1}{2} e^{-2 i \phi } \left(\cos ^2 \theta
                                 \left(4 d+e^{i \chi } z\right) \right) \\ &- \frac{1}{2} e^{-2 i \phi } \left( z \sin \theta
                                 e^{i (\chi +\phi )} \left(2 \sqrt{3} \cos
                                \theta +e^{i \phi } \sin \theta
                                \right)\right). 
\end{align}
These relations give us the three neutrino masses and two Majorana phases from
 Eqs. (\ref{eq:masses}, \ref{eq:phases}).

 Current neutrino experiments cannot observe the three neutrino masses directly. $\beta$-
 decay experiments \cite{Drexlin2013} are sensitive to the effective neutrino mass $m_\beta$ given
 as

\begin{equation}
  \label{eq:mb}
m^2_{\beta} = m^2_1 |U_{e1}|^2 + m^2_2 |U_{e2}|^2 + m^2_3 |U_{e3}|^2.
\end{equation}
The effective neutrino mass $m_{\beta \beta}$ given as
\begin{equation}
  \label{eq:mbb}
m_{\beta \beta} = |m_1 U^2_{e1} + m_2 U^2_{e2} + m_3 U^2_{e3}|.
\end{equation}
can be measured in the neutrino-less double $\beta$-decay experiments \cite{Rodejohann2012a}.
 
 We can obtain the parameters $r,s,$ and $a$ defined in
   Eq. \eqref{eq:kingangles} in terms of $\theta$ and $\phi$ by comparing the
   values of mixing angles given in Eq. (\ref{eq:tm2angles}-\ref{eq:tm2angles3}) with Eq.
   \eqref{eq:kingangles}. These relations are given as follows
\begin{eqnarray}
\label{eq:3}
r & = & \frac{2}{\sqrt{3}}\sin \theta,\\
s &=& \frac{1}{\sqrt{1-\frac{2}{3}\sin^{2}\theta}} - 1,\\
  a &=& \sqrt{1+ \left(\frac{ \sqrt{3} \sin 2\theta \cos \phi }{3-2 \sin ^2\theta }\right)}-1. \\
\end{eqnarray}
 
We perform a Monte Carlo analysis for these two textures by generating the variables $a$,
$d$, $z$ and $\chi$ using uniform random distributions. The parameter space of these
variables is restricted by imposing the experimental constraints \cite{Esteban2017} on $\Delta m^2_{12}=m^2_2-m^2_1$, $\Delta m^2_{23}=m^2_3-m^2_2$,
$\theta_{12}$ and $\theta_{23}$ at 3$\sigma$ C. L. These ranges are shown in Tab.
\ref{tab:experimental}.
The allowed parameter space for $a$, $d$, $z$ and $\chi$ is displayed in Fig. \ref{fig:massvschi} and the allowed ranges can
be read from Tab. \ref{tab:AllowedRanges}. The correlation plots between
$\theta_{23}$ and $\delta$ are presented in Fig. \ref{fig:th23delta} for both the textures. Here, $\theta_{23}$ is well within the experimental range
at $3\sigma$
($[38.4 , 52.8]$, shown as horizontal dashed lines) for
both the textures. However, for 1 $\sigma$ range of
  $\theta_{23}$, we find that only allowed ranges for $\delta$
  are $[215-251]$ and $[287-312]$ ruling out $\delta =[250-286]$.
At $3\sigma$, for both the textures, $\delta$ should be either $90^o$ or
$270^o$ for a maximal $\theta_{23}$. The value of $\delta$ shifts towards $0^o$
or $180^o$ when $\theta_{23}$ takes its extreme values
around $38^o$ or $51^o$. This feature is testable at the experiments like
NO$\nu$A \cite{NOVA_2016_delta_PhysRevLett.116.151806}
and T2K \cite{T2K_NON_ZERO_Theta13_PhysRevLett.107.041801}. The degeneracy in
the Fig. \ref{fig:th23delta} is because of the contributions of positive and negative values of the parameters $a$ and $d$. The allowed values of these parameters can be positive or negative and contribute to the different branches of this figure.

\begin{figure}[h]
  \centering
 \includegraphics[width=0.49\textwidth]{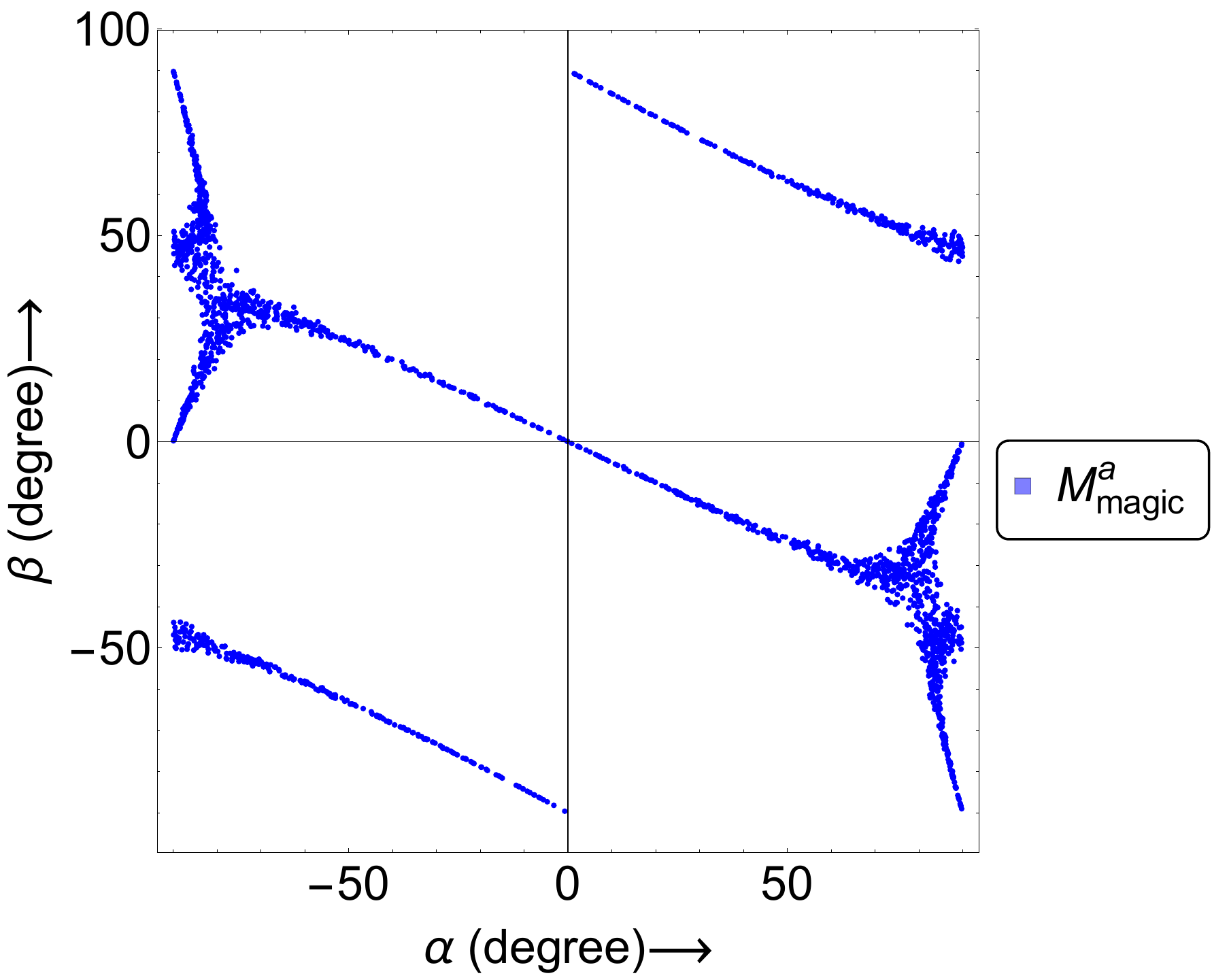}  \includegraphics[width=0.49\textwidth]{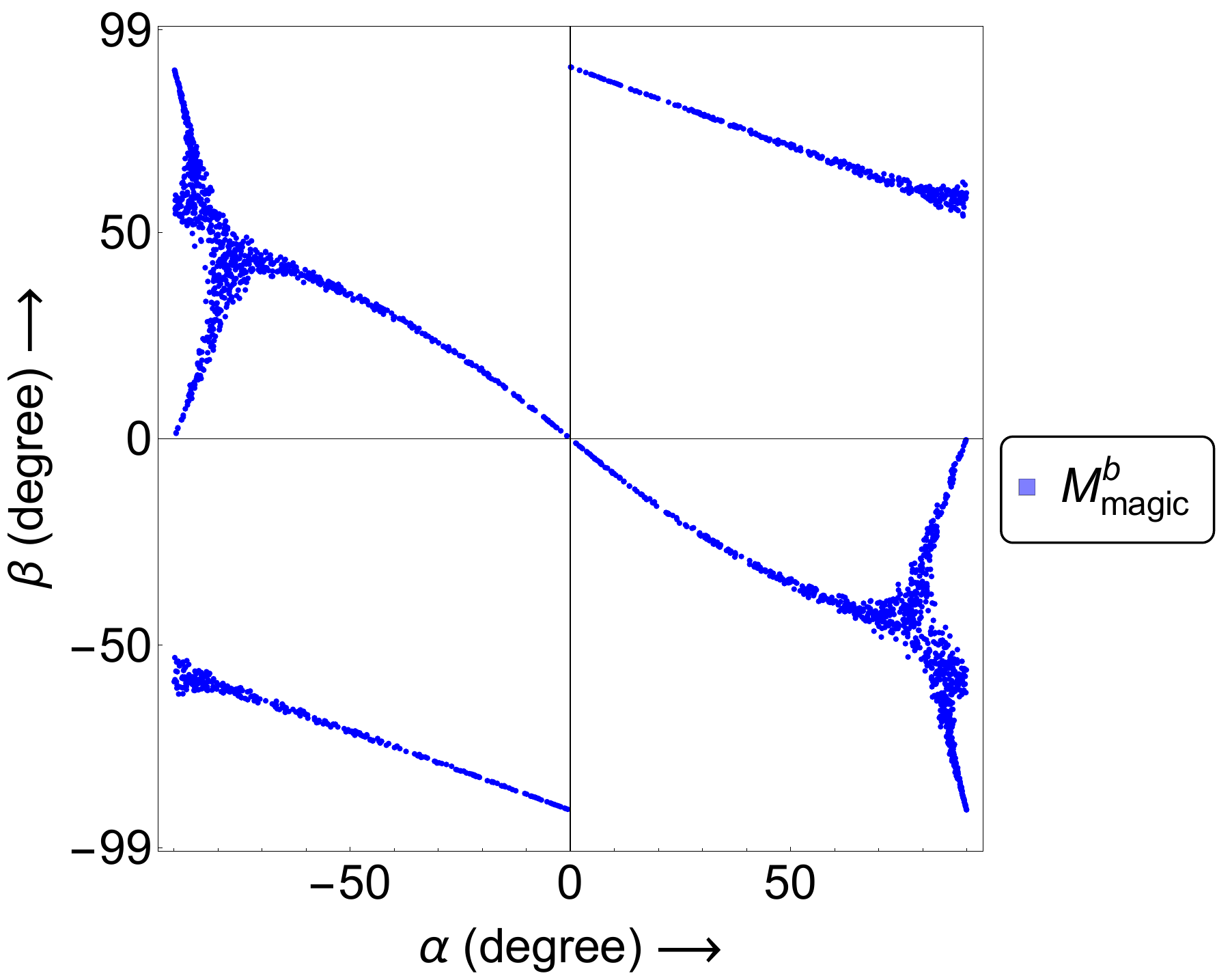}
 \caption{Correlation between majorana phases $\alpha$ and $\beta$ for both
      the textures $M^a_{\text{magic}}$ and $M^b_{\text{magic}}$.}
 \label{fig:alphabeta}
\end{figure}

\begin{figure}[h]
  \centering
\includegraphics[width=0.49\textwidth]{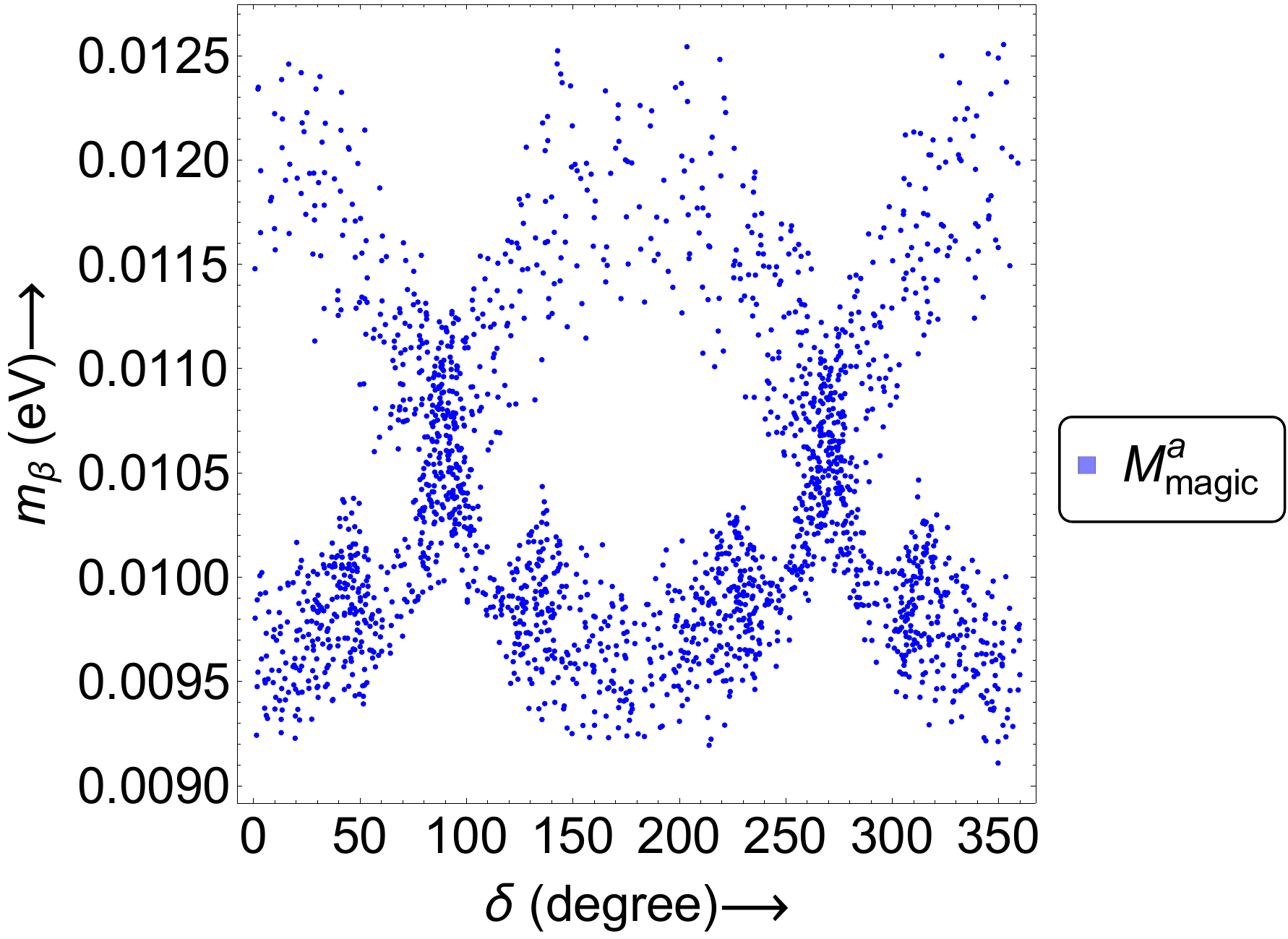} \includegraphics[width=0.49\textwidth]{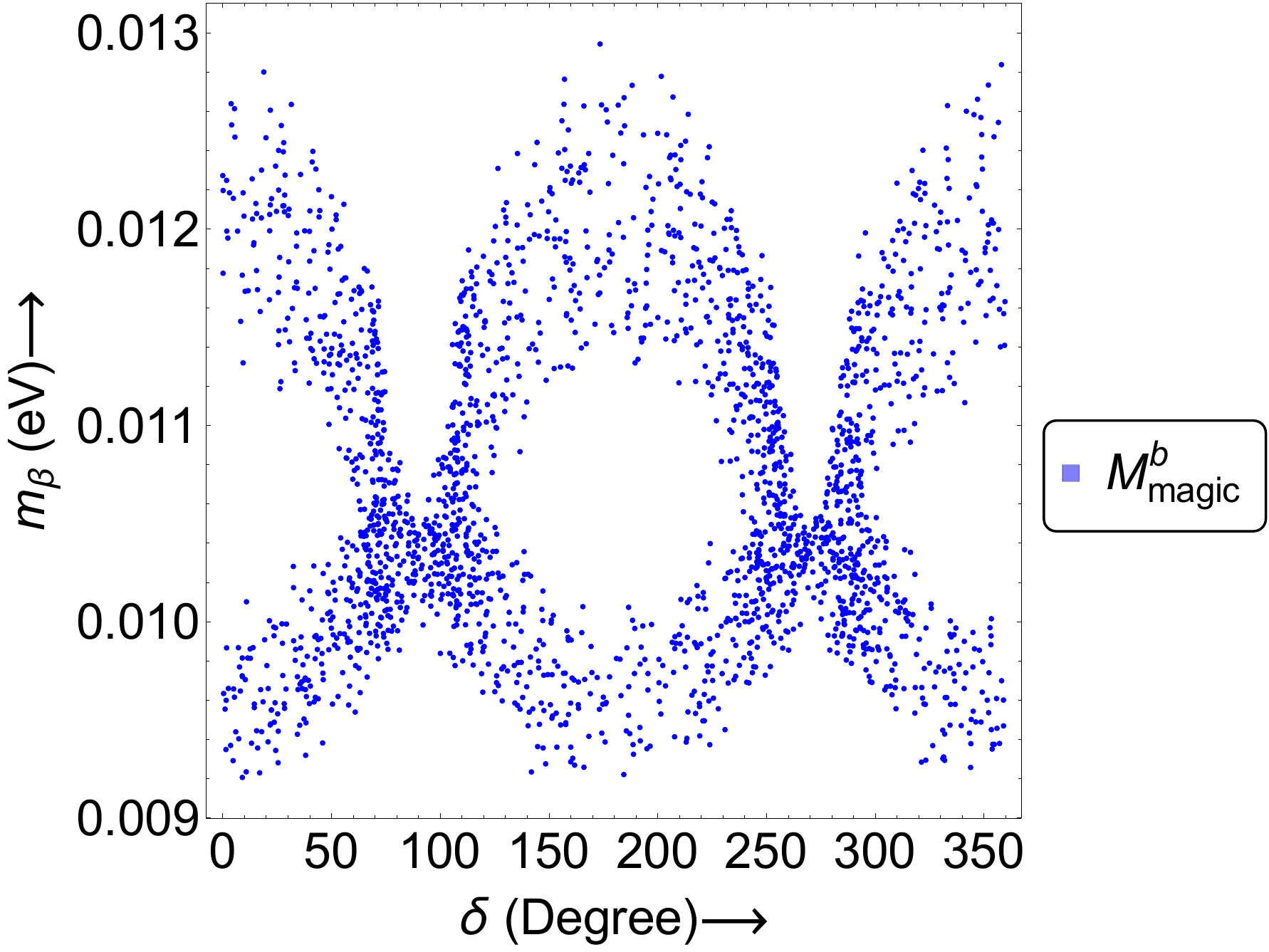}  
\caption{Variation of $m_{\beta}$ with $\delta$ for both
      the textures $M^a_{\text{magic}}$ and $M^b_{\text{magic}}$.}
 \label{fig:mbdelta}
\end{figure}

\begin{figure}[h]
   \centering
\includegraphics[width=0.49\textwidth]{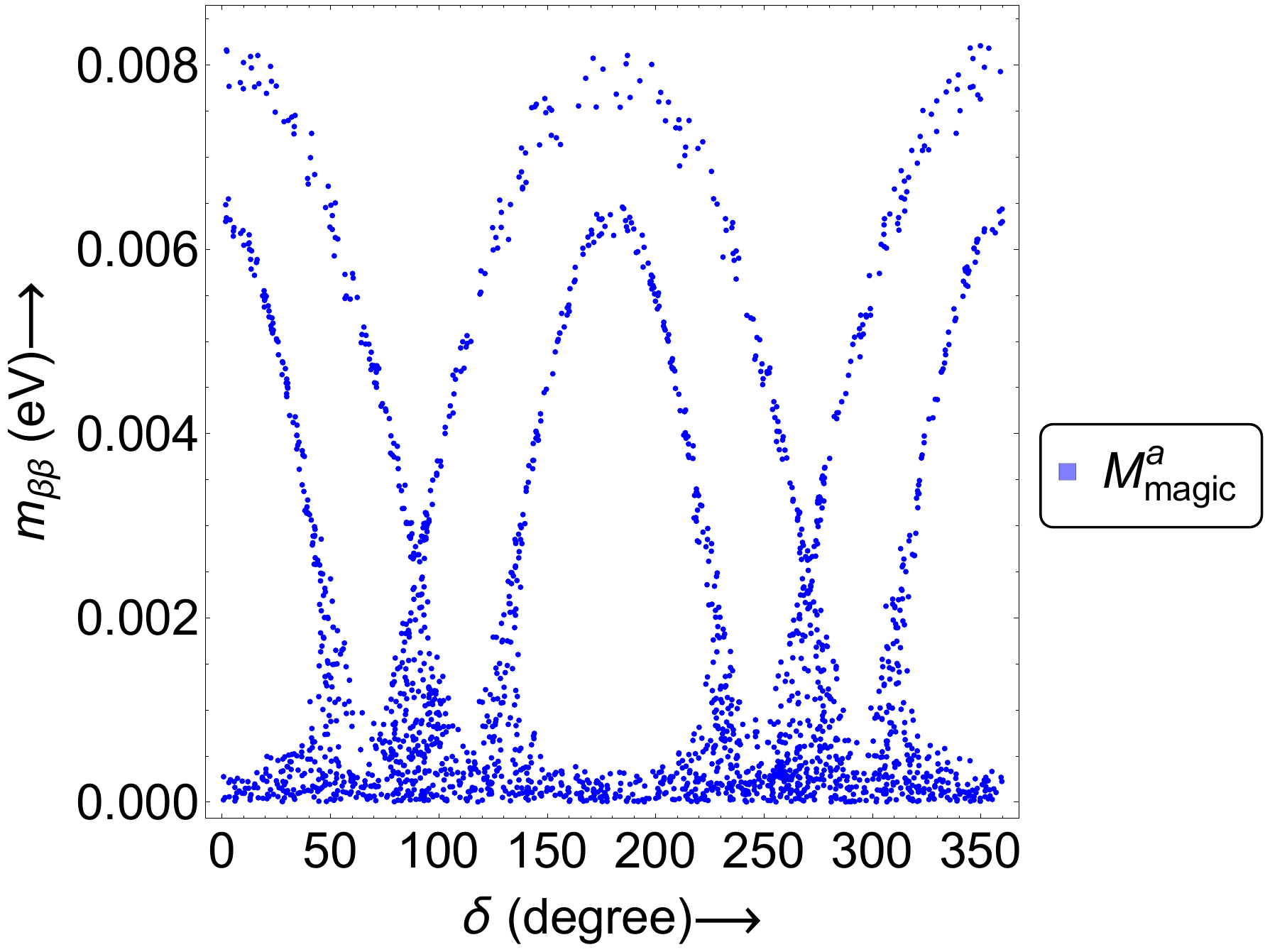} \includegraphics[width=0.49\textwidth]{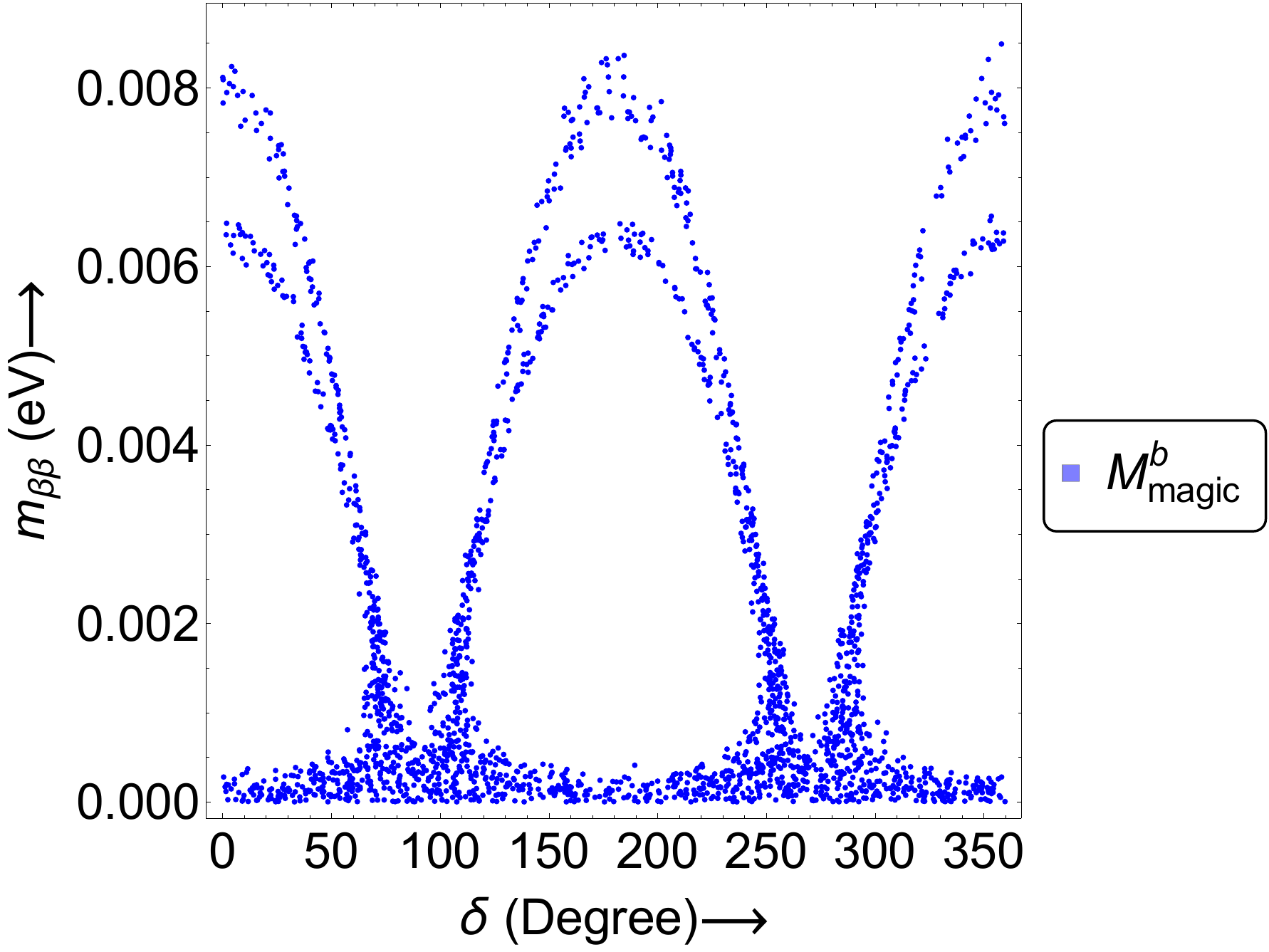}  
\caption{Variation of $m_{\beta \beta}$ with $\delta$ for both
      the textures $M^a_{\text{magic}}$ and $M^b_{\text{magic}}$.}
 \label{fig:mbbdelta}
\end{figure}

The correlation between the phases $\delta$ and $\beta$ is displayed in Fig. \ref{fig:betadelta} and
that between $\alpha$ and $\beta$ is shown in Fig. \ref{fig:alphabeta}. Variation of $m_{\beta}$ and
$m_{\beta \beta}$ with CP violating phase $\delta$ is presented in Fig.
\ref{fig:mbdelta} and \ref{fig:mbbdelta} respectively. 

\begin{figure}[h]
  \centering
\includegraphics[width=0.49\textwidth]{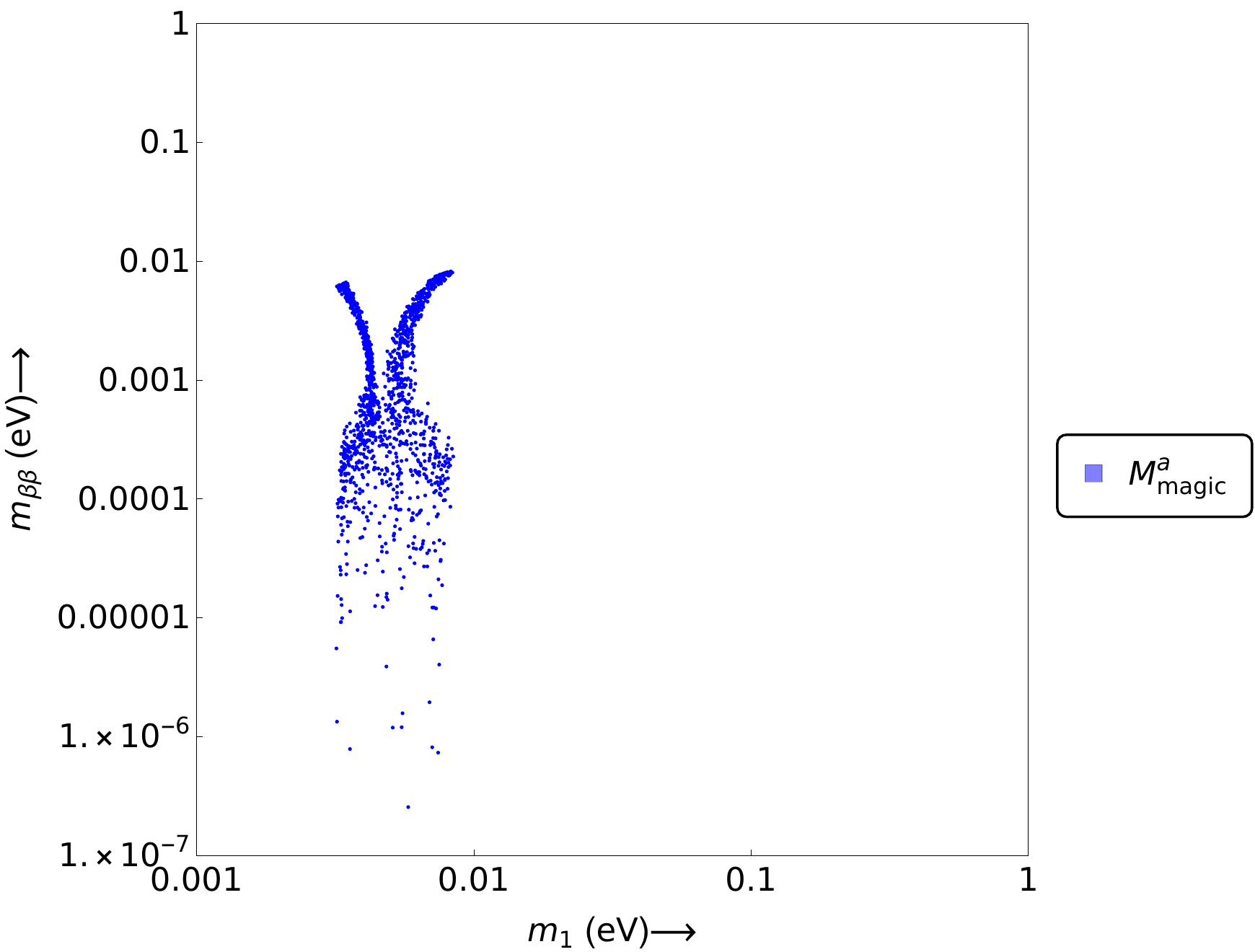} \includegraphics[width=0.49\textwidth]{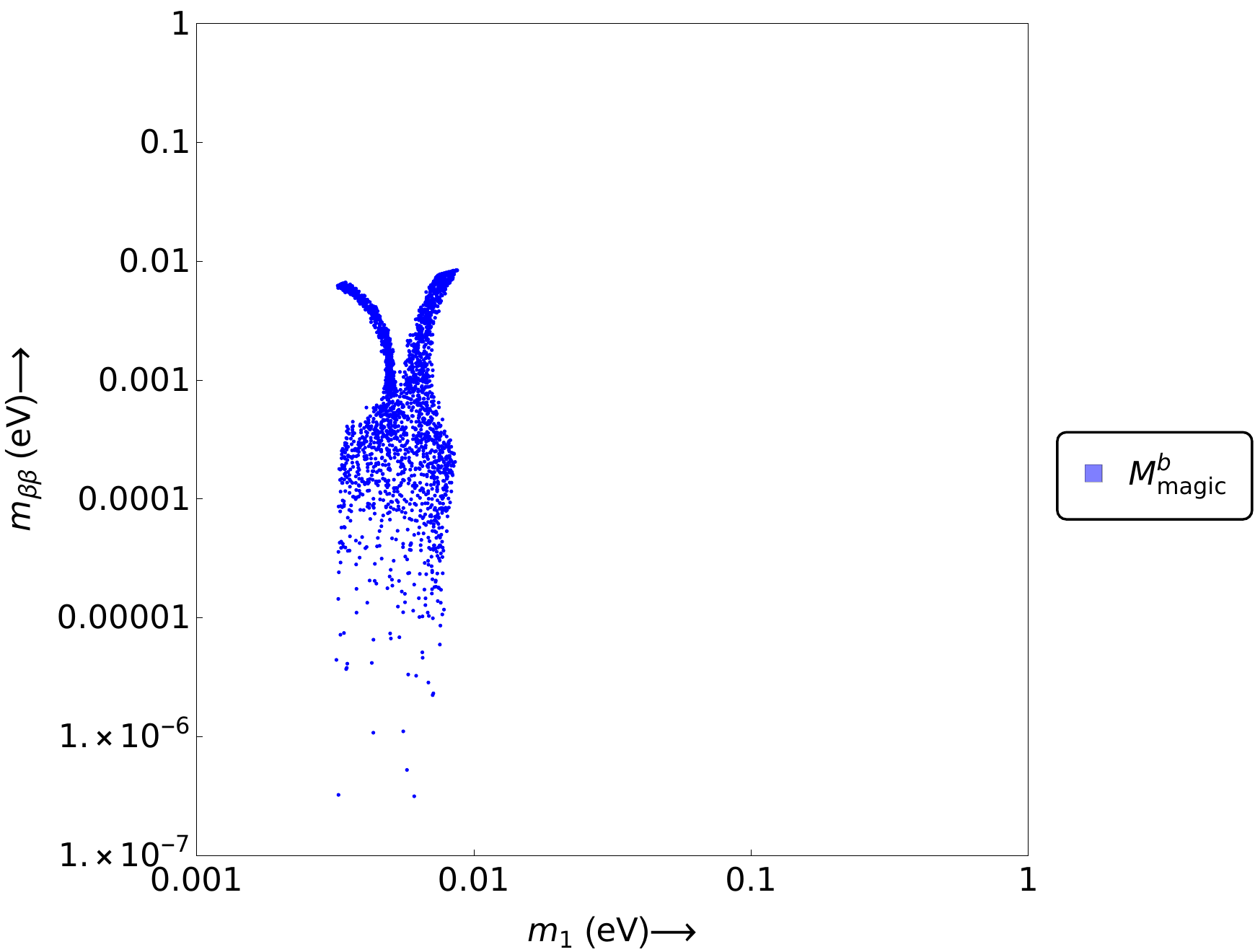}  
\caption{Variation of $m_{\beta \beta}$ with $m_1$ for both
      the textures $M^a_{\text{magic}}$ and $M^b_{\text{magic}}$.}
 \label{fig:mbbm1}
\end{figure}

\begin{figure*}[h]                                                                  
  \centering                                                                       
  \includegraphics[width=0.49\textwidth]{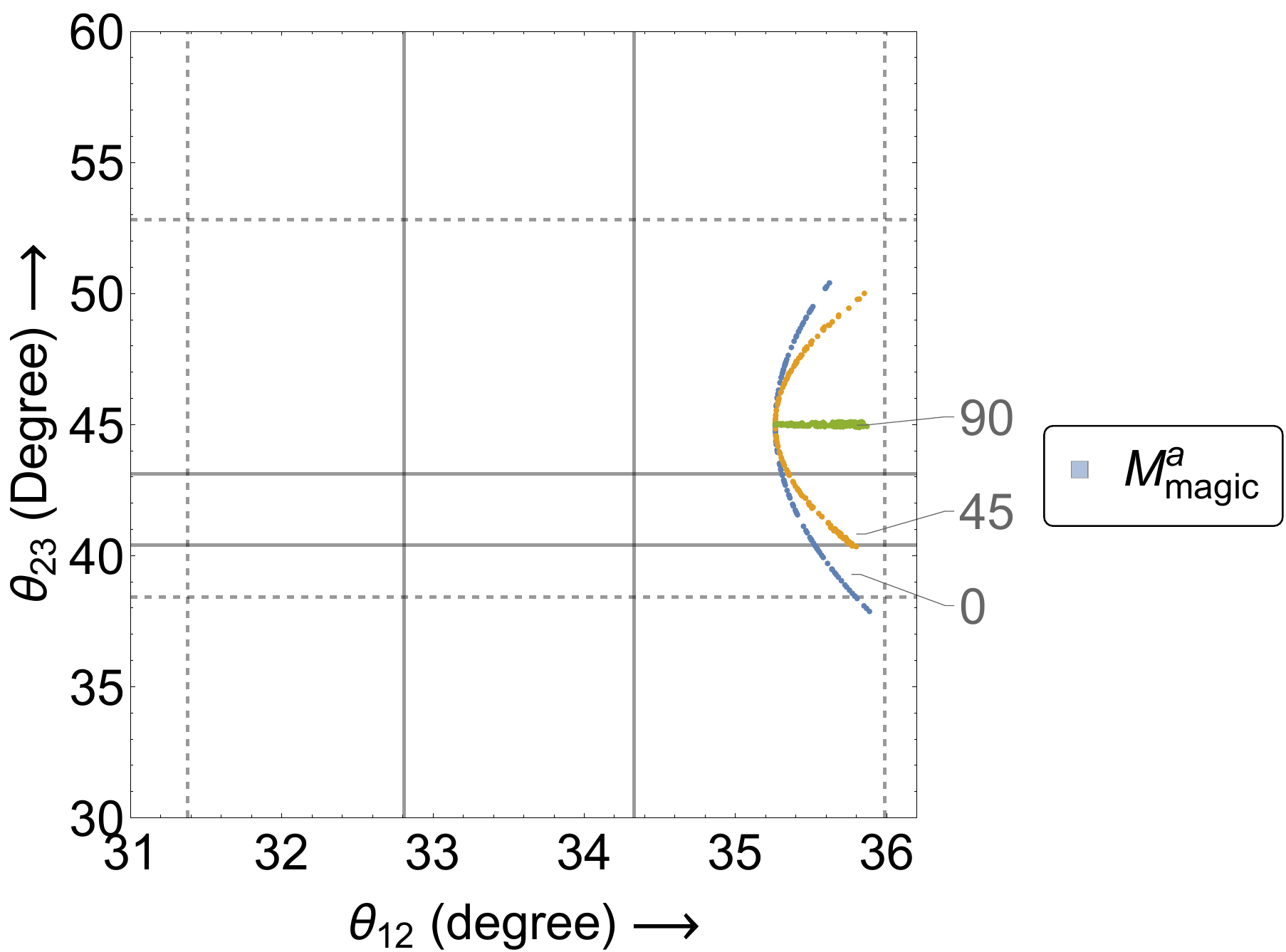}                     
  \includegraphics[width=0.49\textwidth]{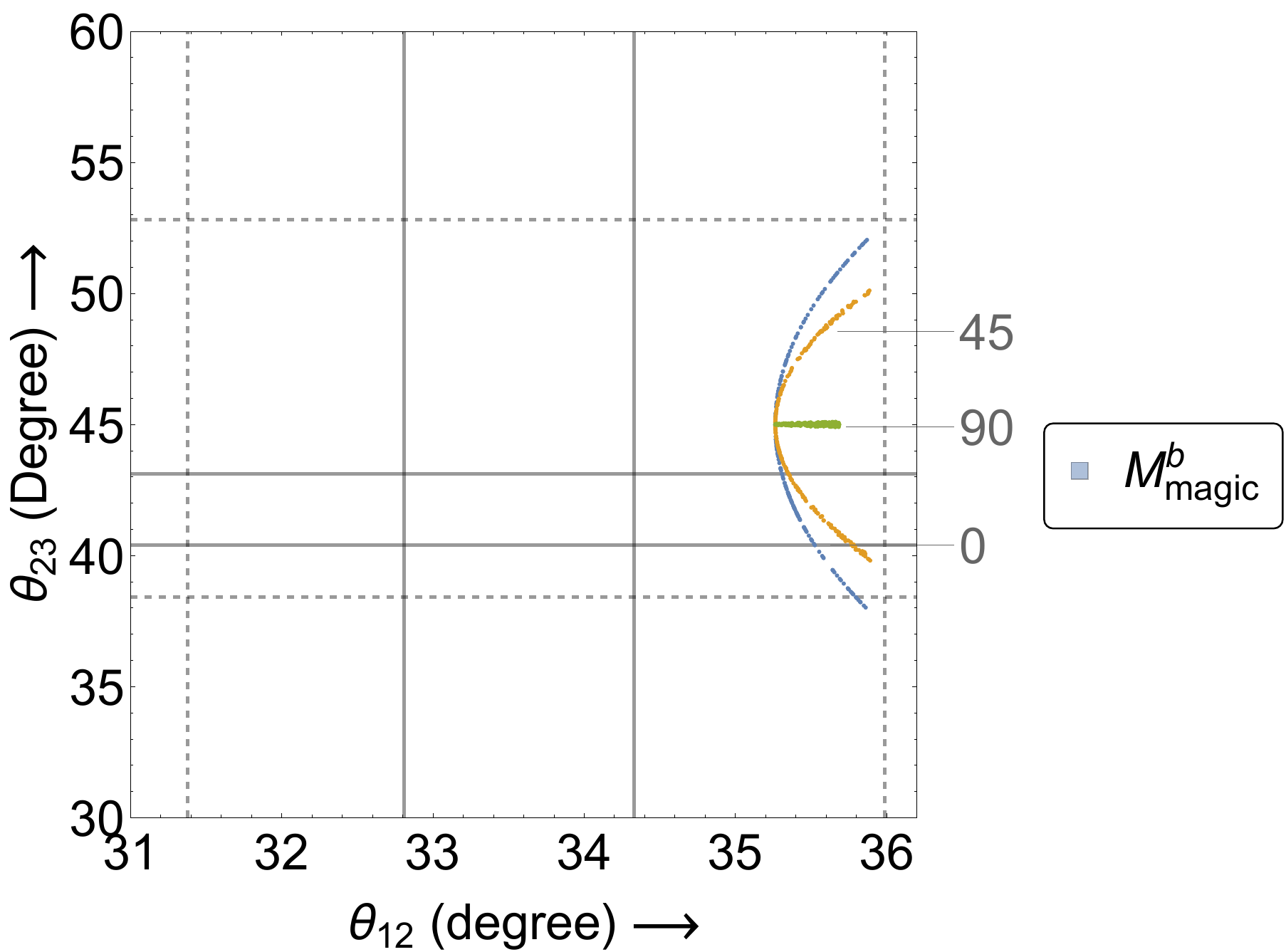}\\                  
  \includegraphics[width=0.49\textwidth]{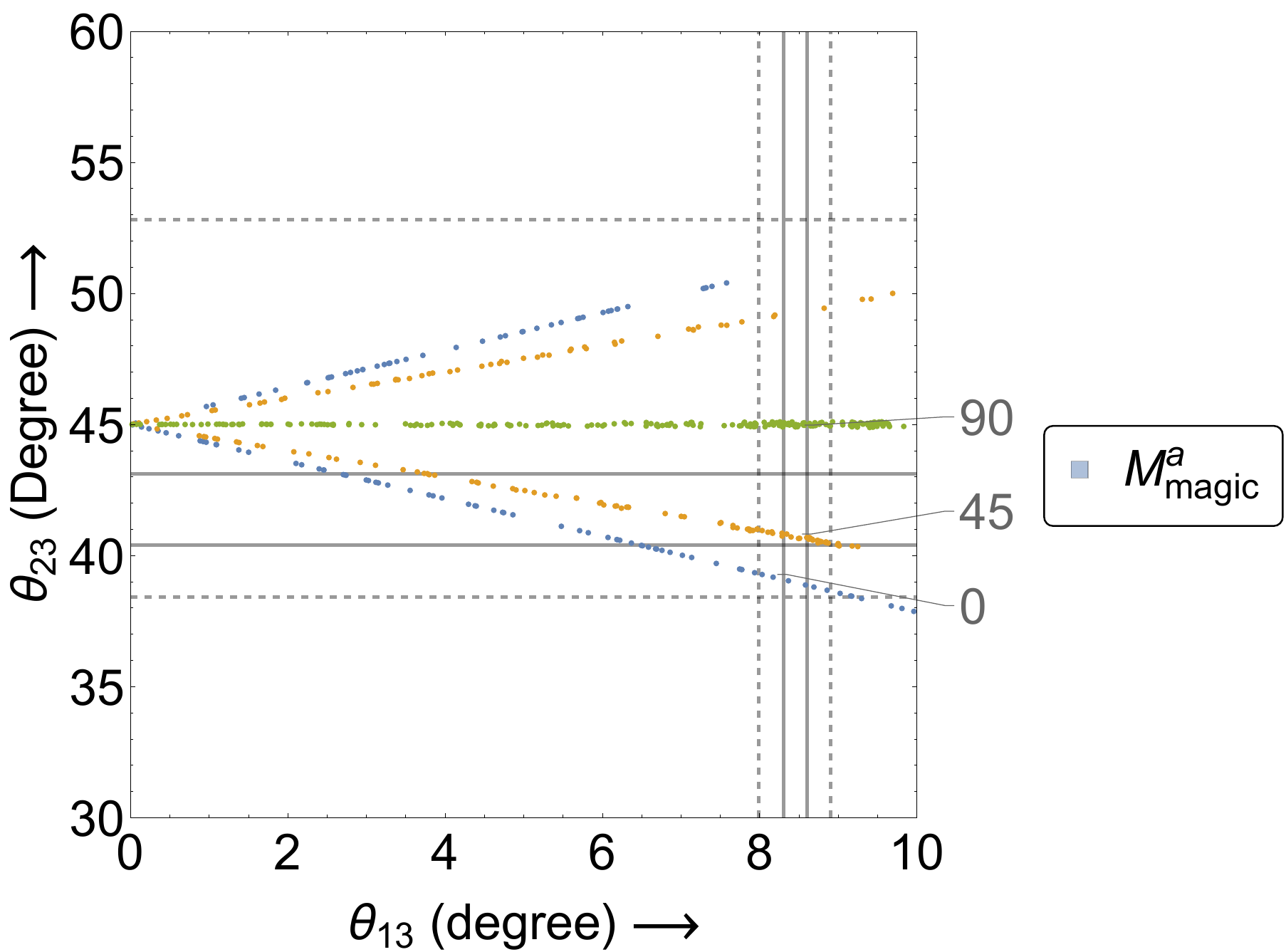}                     
  \includegraphics[width=0.49\textwidth]{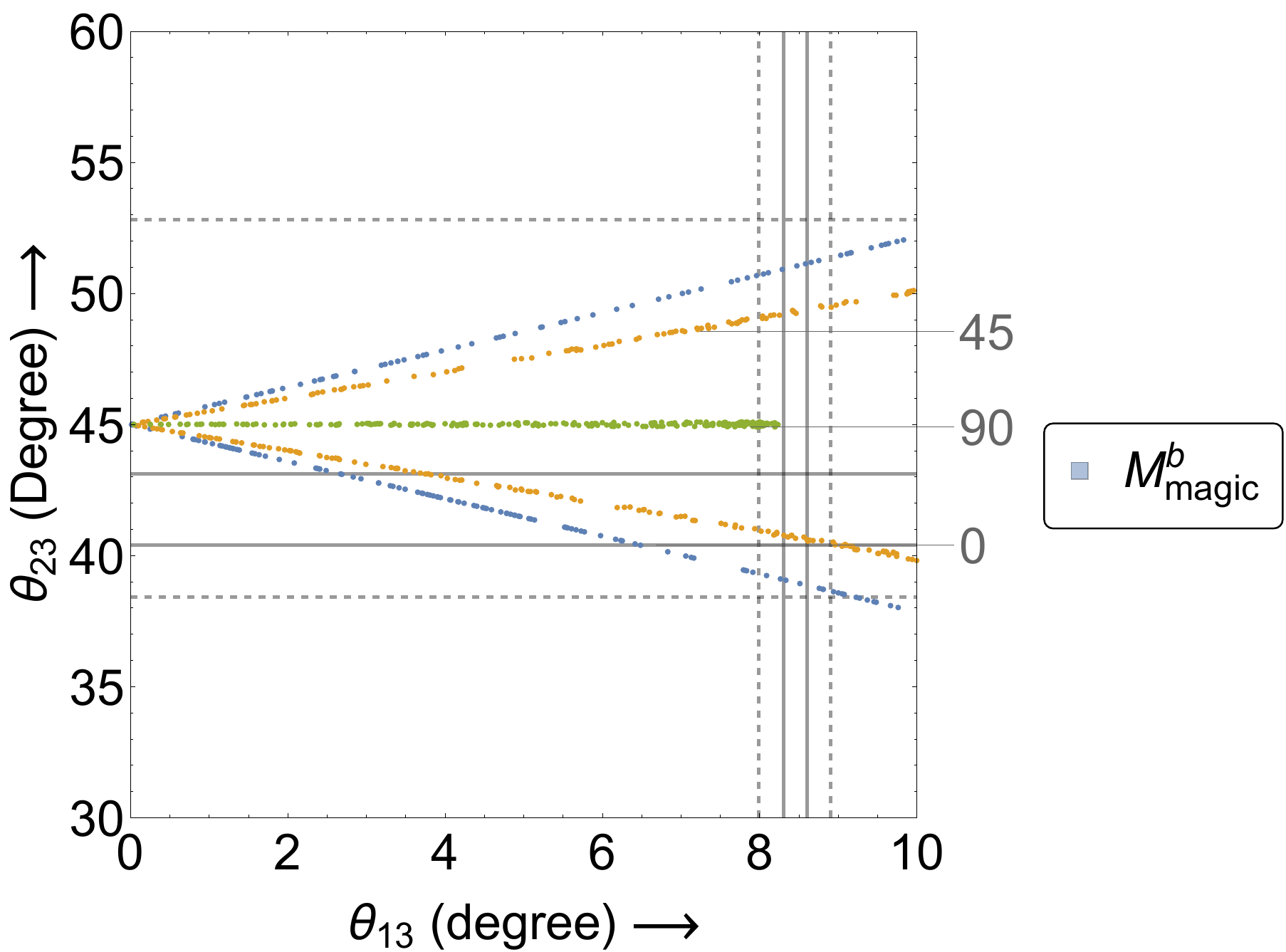}                    
  \caption{Corrections between the mixing angles corresponding a fixed value of
    $\delta$ for both $M^a_{\text{magic}}$ (first column) and
    $M^b_{\text{magic}}$ (second column). Here solid lines represent the
    $1\sigma$ experimental range and dashed lines represent the $3\sigma$
    experimental range.}
  \label{fig:fixeddelta}
\end{figure*}

From Fig. \ref{fig:mbdelta}, \ref{fig:mbbdelta} and \ref{fig:mbbm1}, it is
  clear that our predictions for $m_{\beta}$ and $m_{\beta \beta}$ are very small as compared to
  the sensitivities of the near future $\beta$-decay experiments like KATRIN
  \cite{Drexlin2013,Arenz2018}, Project 8 \cite{Esfahani2017} and double
  $\beta$-decay experiment EXO-200 \cite{Albert2016}, KamLAND-Zen \cite{Gando2016a}. If any of these experiments will be successful in measuring the
  $m_{\beta}$ or $m_{\beta \beta}$, our textures will be ruled out. Correlations between the mixing angles for
  fixed values of $\delta$ (0,45,90,135,180,225,270,315,360) are
  given in the Fig. \ref{fig:fixeddelta}. Here the plots for $(\delta,\delta+180)$
  are identical. Similarly, the plots for
  $(\delta+45,\delta+135,\delta+225,\delta+315)$ and $(\delta+90,\delta+270)$
  are same. The solid lines represent the
$1\sigma$ experimental range and dashed lines represents the $3\sigma$
experimental range.

\begin{figure}[h]
  \centering
\includegraphics[width=0.49\textwidth]{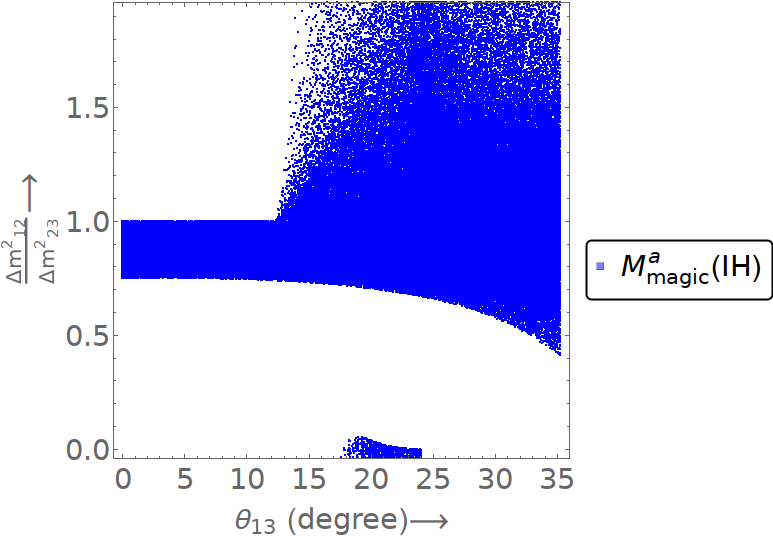} \includegraphics[width=0.49\textwidth]{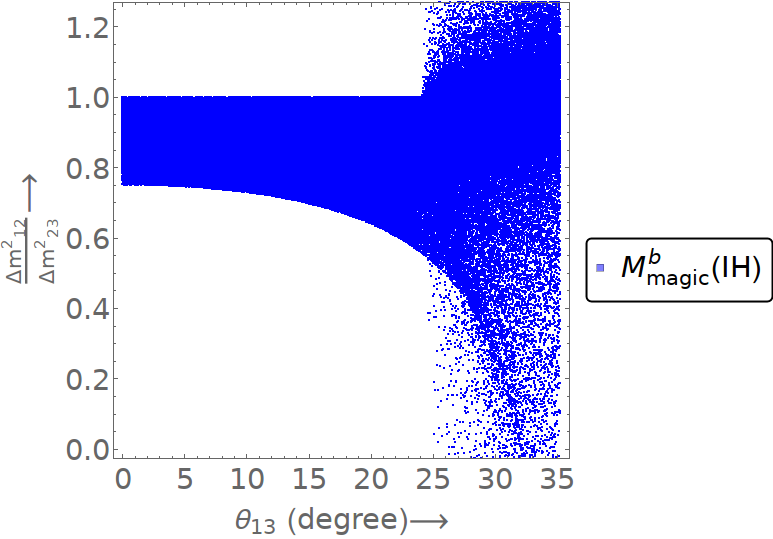}  
\caption{Variation of $\frac{\Delta m^2_{21}}{\Delta m^2_{23}}$ with $\theta_{13}$ for both
      the textures $M^a_{\text{magic}}$ and $M^b_{\text{magic}}$ assuming
      inverted hierarchy of neutrino masses.}
 \label{fig:ih}
  \end{figure}

We have constructed the textures $M^a_{\text{magic}}$ and $M^b_{\text{magic}}$
for normal hierarchy with vanishing lowest eigenvalue of $M_{TBM}$. We can
obtain similar textures for neutrino masses with inverted hierarchy which can
also be written as $M^{i}_{\text{magic}} = M_{\text{TBM}}+M'_i$. Assuming that
$M_\text{TBM}$ has lowest vanishing eigen value, the
forms of $M_\text{TBM}$ and $M'_i$ for the inverted hierarchy case are as
follows:
\begin{equation}
  \label{eq:inverted hierarchy}
M_{\text{TBM}} = \left(
\begin{array}{ccc}
 a & a+2 d & a+2 d \\
 a+2 d & a+d & a+d \\
 a+2 d & a+d & a+d \\
\end{array}
\right), M'_a = \left(
\begin{array}{ccc}
 0 & 0 & \eta  \\
 0 & 0 & \eta  \\
 \eta  & \eta  & -\eta  \\
\end{array}
\right), M'_b = \left(
\begin{array}{ccc}
 0 & \eta  & 0 \\
 \eta  & 0 & 0 \\
 0 & 0 & \eta  \\
\end{array}
\right).
\end{equation}
However, as shown in the Fig. \ref{fig:ih}, we find that the experimental ranges for the mixing angle $\theta_{13}$ and the
ratio $\Delta m^2_{12} / |\Delta m^2_{23}|$ cannot be satisfied simultaneously
for the inverted hierarchy case. Thus, the corresponding textures with inverted hierarchy are ruled out.

In one of our previous study \cite{Channey2017}, we had demonstrated the idea that viable textures $TM_{1}$ and $TM_{2}$ of the neutrino mass matrix can be created which are modifications of the mass matrix corresponding to TBM mixing. We did not provide any rationale to these textures. However in the present work, we propose a systematic method to modify the neutrino mass matrix corresponding to the TBM mixing $M_{TBM}$. We modify the $M_{TBM}$ by breaking $\mu - \tau$ symmetry but preserving the magic symmetry. We generate two such textures $M^{a}_{magic}$ and $M^{b}_{magic}$. The texture $M_{2}$ of our previous study and the texture $M^{b}_{magic}$ of our present study are related by $\mu - \tau$ exchange symmetry resulting in identical predictions for these textures. The another texture $M_{1}$ of the previous study did not preserve either $\mu - \tau$ symmetry or the magic symmetry. Similarly the texture $M^{a}_{magic}$ of our current study is different from any of the previously proposed textures as can be seen by comparing the Fig. \ref{fig:massvschi}, \ref{fig:mbdelta}, \ref{fig:mbbdelta} and \ref{fig:mbbm1} of the present manuscript.

In conclusion, we have investigated two simple textures of the neutrino
mass matrix with magic symmetry. These textures can be written as combination of
TBM mass matrix with a vanishing eigenvalue and a simple perturbation matrix with
one complex parameter preserving the magic symmetry. These textures have four
real parameters: $a$, $d$,  $z$, and $\chi$. We find the allowed ranges for
these parameters and present the resulting correlations between $\theta_{23}$
and $\delta$. We find that $\delta$ should be around $90^o$ or $270^o$ for
maximal $\theta_{23}$ mixing for both textures. When $\theta_{23}$ is around
$40^o$ or $50^o$, $\delta$ can take values near $0^o$ or $180^o$. Such
correlations are generic features of magic symmetry and are testable at future
neutrino experiments like NO$\nu$A and T2K. Our textures have definite
predictions for $m_\beta$
\cite{Drexlin2013,Arenz2018,Esfahani2017,Abdurashitov2015,Fraenkle:2015yfs} and
$m_{\beta \beta}$ \cite{Gando2016,Ebert2016,Agostini2017} which can be tested at $\beta$-decay and neutrino-less double $\beta$-decay experiments.

\clearpage
\bibliographystyle{elsarticle-num}
\bibliography{library,first,library2}
\end{document}